\documentclass[11pt]{article}
\usepackage{graphics}
\usepackage{psfrag}
\usepackage{epsfig}
\usepackage{epsf}
\newcommand{\str}{\quad \Rightarrow \quad}
\newcommand{\fiverm}{\small}
\newcommand{\sevenrm}{\small}
\newcommand{\fet}[1]{\mbox{\boldmath $#1$}}
\newcommand{\lapprox}{{\renewcommand{\arraystretch}{0.3}
\begin{array}{c} {\tiny < } \cr {\tiny \sim} \end{array}}}
\newcommand{\beq}{\begin{equation}}
\newcommand{\eeq}{\end{equation}}
\newcommand{\beqa}{\begin{eqnarray}}
\newcommand{\eeqa}{\end{eqnarray}}
\newcommand{\nn}{\nonumber \\ }
\renewcommand{\thefootnote}{\#\arabic{footnote}}
\newcommand{\ve}{\varepsilon}
\newcommand{\krig}[1]{\stackrel{\circ}{#1}}
\newcommand{\barr}[1]{\not\mathrel #1}
\newcommand{\pr}{\overrightarrow}
\newcommand{\lev}{\overleftarrow}
\renewcommand{\arraystretch}{1.3}
\newcommand{\vs}{\vspace{-0.25cm}}

\begin{document}

\hfill FZJ-IKP(TH)-1998-24


\vspace{1in}

\begin{center}

{{\Large\bf Effective theory for the two--nucleon system}}

\end{center}

\vspace{.3in}
\begin{center}
{\large
E.~Epelbaoum,$^{a,b}$\footnote{email:
evgeni.epelbaum@hadron.tp2.ruhr-uni-bochum.de}
W.~Gl\"ockle,$^a$\footnote{email:
walter.gloeckle@hadron.tp2.ruhr-uni-bochum.de}
A.~Kr\"uger,$^a$\footnote{email:
andreas.krueger@hadron.tp2.ruhr-uni-bochum.de}
Ulf-G. Mei{\ss}ner$^b$\footnote{email: Ulf-G.Meissner@fz-juelich.de}}

\bigskip

$^a${\it Ruhr-Universit\"at Bochum, Institut f{\"u}r
  Theoretische Physik II\\ D-44870 Bochum, Germany}

\bigskip

$^b${\it Forschungszentrum J\"ulich, Institut f\"ur Kernphysik
(Theorie)\\ D-52425 J\"ulich, Germany}

\end{center}

\vspace{.9in}
\thispagestyle{empty}

\begin{abstract}\noindent
We apply the method of unitary transformations to a model two--nucleon
potential and construct from it an effective potential in a subspace of momenta
below a given cut--off $\Lambda$. The S--matrices in the
full space and in the subspace are shown to be identical.
We solve numerically
the Schr\"odinger equation in the small momentum space and recover exactly the
bound and scattering states of the full theory. We then expand the heavy
repulsive meson exchange of the effective potential in a series of local
contact terms and discuss the question of naturalness of the corresponding
coupling constants.  Using our exact effective theory we address
further issues related to the chiral perturbation theory approach of the
two--nucleon
system. The coordinate space representation of the effective potential
is also considered.
\end{abstract}

\vspace{2cm}

\begin{center}

{\it Accepted for publication in Nuclear Physics~A}

\end{center}
\vfill

\pagebreak

\section{Introduction}
\def\theequation{\arabic{section}.\arabic{equation}}
\setcounter{equation}{0}
Chiral perturbation theory is hoped to provide insight into the nuclear force
problem and possibly even  lead to  a quantitative framework. Chiral symmetry
imposes constraints on possible momentum and spin dependences of the nuclear
forces but the framework is restricted to  momenta below a certain
scale. Only in this regime one can set up a  power counting scheme which
limits the number of interaction terms in the nuclear Hamiltonian. Whether
this scheme works quantitatively and is applicable to interacting nucleons
seems not yet fully established, despite a large number of investigations
and considerable progress over the last few years
(an incomplete list of references
is~\cite{weinberg}--\cite{kor}.
For a short state--of--the--art review see e.g.~\cite{silas2}). We therefore
think that a model study based on a simplified nuclear force which, however,
captures the essential features  of the nucleon--nucleon interaction
(long/intermediate--range attraction and short--range repulsion)
can provide useful insights.

In this study, we investigate the two--nucleon
system in a space of momenta whose values are below a given cut--off
$\Lambda$. For that, we start with a model nucleon--nucleon (NN) interaction
given in full momentum space. The potential consists of two terms, an
attractive one due to the exchange of a light meson with mass $\mu_L$ and
a repulsive term parametrized in terms of a heavy meson exchange with
a mass $\mu_H$. The strengths parameters accompanying these terms are
determined by fitting physical properties like e.g. the deuteron binding
energy.
We then divide the momentum space into two
subspaces, spanning the values from zero to $\Lambda$ and from $\Lambda$ to
$\infty$, respectively.
Consequently, the two--nucleon Hamiltonian can be regarded as a two--by--two
matrix
connecting the two momentum subspaces. By an unitary transformation it can
be block--diagonalized decoupling the two subspaces. In this manner one can
construct an effective Hamiltonian acting only in the low momentum
subspace. The so constructed effective Hamiltonian
comprises the full physics for
low--lying bound  and scattering states with
appropriate boundary conditions. Specifically, for the scattering states
the initial momenta should belong to the low momentum
subspace. Nevertheless, and this is an important remark,
the physics of the high momenta is by the very construction
included in the effective low momentum theory. Like in treatments
performed in the context of chiral perturbation theory all or some of the
resulting effective interaction can be cast into the form of a string of
contact interactions of increasing powers in the momenta. For a consistent
power counting to emerge, the coefficients
accompanying these terms should be of  natural size. In other
words there should be a momentum scale (the naturalness scale
$\Lambda_{\rm scale}$) such that the properly normalized
coefficients are of order one. In our exact effective theory, we can
in fact precisely calculate these coefficients and check the naturalness
property. Stated differently, the effective theory\footnote{We stress
that the effective theory considered here is not an effective field
theory (EFT). This important distinction should be kept in mind.}
obtained by the exact
momentum space projection plays the role of QCD and the expansion in
terms of contact interactions for the heavy meson exchange is our model
of the effective field theory. Note that throughout this investigation,
we keep the light meson (pion) exchange explicitly.
Further questions that  can be addressed are: a)  can
one find a subdivision of the full momentum  range such that the
effective interaction is only weakly dependent on the
precise choice of the value $\Lambda$, and b) is there relation between
$\Lambda$ and the natural momentum scale? We can also investigate the
convergence properties of the expansion in terms of local operators since
we have the exact solution to the problem under control.
The effective interaction is naturally constructed in momentum space,
but one can consider it in configuration space as well. Clearly,
one has to expect the effective potential to look
totally different compared to the original potential and as a
consequence,  the deuteron wave function will also change. More precisely,
the projection of the original potential into the subspace of small momenta
induces non--localities in momentum space which in turn lead to very
complicated co--ordinate space expressions. We will show some instructive
examples of this phenomenon.

In section II we formulate the model and the way the block--diagonalization
is performed. We use an unitary operator proposed long time ago by
Okubo~\cite{okubo} (see also ref.\cite{FST}),
which leads to a nonlinear decoupling equation.
We show that the T--matrix evaluated with the help
of the effective interaction is {\it exactly} equal to the original T--matrix
for the underlying potential.  This is related to the fact that the
component of the transformed wave function in the
subspace of the high momenta above $\Lambda$ is exactly zero
for the low energy  scattering  states in which we are interested.
This proof is different from the one given in refs.\cite{okubo}\cite{FST}.
An alternative way, which, however, assumes the knowledge of the the
scattering states, leads
to a linear equation~\cite{SO}. 
This is briefly discussed in the end of the section.

The numerical investigations are described in section III. It is shown how
the nonlinear decoupling equation can be solved by a suitably chosen iteration.
For that, one has to
modify the original potential in order to avoid difficulties
when the momentum exactly equals the chosen cut--off value.
Physical observables, however, do not depend on this modification.
We demonstrate  (numerically) the exact agreement for the binding energy
and scattering phase shifts using the effective and the original
Hamiltonian, respectively. We then expand the heavy meson exchange term
in a series of local contact terms with increasing dimension. This allows us
to extract  information on the naturalness property
of the expansion coefficients in the effective interaction. As an application
we calculate the expectation values of the various terms in the effective
potential with respect to the effective deuteron wave function  and low energy
scattering states. This allows to study the convergence properties of the
contact term
expansion. We discuss the relevance of these results for a systematic
effective field theory treatment of the two--nucleon system.
We also regard
the configuration space representation of the effective interaction.
Note that some of these results were already  discussed in the
letter~\cite{egmplb}.

We summarize and conclude in section IV.

\section{Projection formalism for the effective theory}
\setcounter{equation}{0}

In this section we develop in detail the formalism which allows to
study the nucleon--nucleon interaction in a Hilbert space of momenta
below a chosen cut--off in momentum space. The starting point is a
given potential which involves all momentum scales and reproduces
qualitatively the two--nucleon phenomenology (in the S-waves). Besides
being interesting in itself, such an
effective theory in a space of low momenta can also be used
to study  various aspects of effective field theory approaches to the
nucleon--nucleon interaction. For simplicity, we restrict ourselves
to the S--waves. The formalism is, however, more general and can
straightforwardly be extended to more complicated potentials.

To be specific, consider a momentum space Hamiltonian for the
two--nucleon system of the form
\beq
\label{ham}
H (\vec{p}, \vec{p\,}') = H_0 ( \vec{p}\,) \delta (\vec{p} - \vec{p\,}')
+ V ( \vec{p}, \vec{p\,}')~,
\eeq
where $H_0$ stands for the kinetic energy and $V$ for the spin--independent
model force. We introduce the projection operators
\beqa
\label{eta}
\eta &=& \int d^3 p \left| \vec{p}\, \right\rangle \left\langle \vec{p}\,
  \right| \,\, ,  \quad \quad \quad
\left| \vec{p} \, \right| \leq \Lambda \,\, , \\
\label{lambda}
\lambda &=& \int d^3 p \left| \vec{p} \,\right\rangle  \left\langle
\vec{p} \, \right| \,\,  ,
\quad \quad \quad
\left| \vec{p} \, \right| > \Lambda \,\, ,
\eeqa
where $\Lambda$ is a momentum cut--off which  separates the low from the high
momentum region. Its precise value will be given below.
Apparently $\eta^2 = \eta$, $\lambda^2 = \lambda$,
$\eta \lambda = \lambda \eta =0$ and $\eta + \lambda =1$.
Using  the $\eta$ and $\lambda$ projectors, the Schr\"odinger
equation takes the form
\begin{equation}
\label{2}
\left( \begin{array}{cc} \eta H \eta & \eta H \lambda \\
\lambda H \eta & \lambda  H
\lambda \end{array} \right) \left( \begin{array}{c} \eta | \Psi \rangle \\
\lambda | \Psi \rangle \end{array} \right)
= E  \left( \begin{array}{c} \eta | \Psi \rangle \\
\lambda | \Psi \rangle \end{array} \right)
\quad .
\end{equation}
Obviously, the
low and the high momentum components are coupled. Our aim is to derive an
effective Hamiltonian acting on low momentum states only and
which furthermore incorporates
all the physics related to the possible bound  and scattering states
with initial asymptotic momenta from the  $\eta$ states. This can be
accomplished by an unitary transformation $U$
\beq
\label{transf}
H \longrightarrow H ' = U^\dagger H U
\eeq
such that
\beq
\label{bed}
\eta H ' \lambda = \lambda H ' \eta = 0~.
\eeq
We choose a parametrization of $U$ given by Okubo~\cite{okubo},
\begin{equation}
\label{Uokubo}
U = \left( \begin{array}{cc} (1 + A^\dagger A )^{- 1/2} & -
A^\dagger ( 1 + A A^\dagger )^{- 1/2} \\
A ( 1 + A^\dagger A )^{- 1/2} & (1 + A A^\dagger )^{- 1/2} \end{array}
\right)~,
\end{equation}
where $A$ has to satisfy the condition
\begin{equation}
\label{condA}
A = \lambda A \eta~.
\end{equation}
It is then straightforward  to recast
the conditions eq.(\ref{bed}) in a different form,
\begin{equation}
\label{Aeq}
\lambda \left( H - \left[ A, \; H \right] - A H A \right) \eta = 0~.
\end{equation}
This is a nonlinear equation for the operator A, which takes the
explicit form
\begin{eqnarray}
\label{eqa}
{V} (\vec{p} ,\vec{q}\, ) &-& \int  d^3q' \,
A( \vec{p}, \vec{q}\, ' ) {V}( \vec{q}\,' , \vec{q}\,)
+ \int  d^3p' \,
{V} (\vec{p} , \vec{p}\,' ) A (\vec{p}\,' , \vec{q}\, )
  \nonumber\\
&-& \int  d^3q' \,  d^3p' \,
A(\vec{p} , \vec{q}\,' ) {V} (\vec{q}\, ' , \vec{p}\, ' )
A(\vec{p}\, ' , \vec{q}\, )
\nonumber\\
&=& (E_{{q}} - E_{{p}}) \, A(\vec{p} , \vec{q} \,) \quad.
\end{eqnarray}
Here we denoted the momenta from the  $\eta$ and $\lambda$--spaces by $\vec{q}$
and $\vec{p}$, respectively,
and $E_q$, $E_p$ stand for the corresponding kinetic energies.
Once $A$ and thus $U$ have been determined, the effective Hamiltonian in the
$\eta$--space takes the form
\begin{equation}
\label{14}
\eta H ' \eta = \eta ( 1 + A^\dagger A)^{-1/2}
\left( H + A^\dagger H + H A + A^\dagger H A
\right) (1 + A^\dagger A)^{-1/2} \eta~.
\end{equation}
This interaction is by its very construction energy--independent
and hermitean~\cite{okubo}\cite{FST}\cite{EGM}.

After this block--diagonalization, the Schr\"odinger equation
separates into two effective equations in the respective  subspaces.
According to  eq.(\ref{transf}) the connection
between the eigenstates of $H$ and $H '$ is
\beq
\label{conn}
\Psi ' = U^\dagger \Psi~,
\eeq
so that  a priori
the transformed problem separates into
\beqa
\eta H ' \eta \Psi ' &=& E \, \eta \Psi ' ~,\label{Seta} \\
\lambda H ' \lambda \Psi ' &=& E \, \lambda \Psi ' ~.
\label{Slam}
\eeqa
However, one of the components of $\Psi '$ has to be identically zero \cite{privet}.
If both $\eta$ and $\lambda$ components of $\Psi '$ would be different from zero, one 
could form any linear combination thereof and would end up in the original
space with an infinite degeneracy, which is incorrect. 
It remains to be clarified in which 
space the transformed scattering and bound states reside. 
Let us first regard scattering states.
In which space lies $\Psi '$ if the scattering process is initiated by    
asymptotic momenta from the low momentum range spanned by $\eta$?
For that, we define the effective potential by
\beq
V '  \equiv H ' - H_0 \quad ,
\eeq
which is, of course, block--diagonal, since $H_0$ and $H '$ have this property.
Note further that the resolvent operator of the transformed
Hamiltonian $H '$ is block--diagonal as well:
\beq
\label{resolv}
U^\dagger ( z - H )^{-1} U = ( z - H ' )^{-1} \equiv
G ' (z)  = \left( \begin{array}{cc} (z - \eta H ' \eta)^{-1} &
0 \\ 0 & (z - \lambda H ' \lambda)^{-1} \end{array} \right)~.
\eeq
Obviously, $z$ should not be in the spectrum of $H$.
This block--diagonal form (\ref{resolv}) has the immediate consequence
that the scattering states $| {\Psi_{\vec q} '}^{(+)} \rangle $,
defined via
\beq
| {\Psi_{\vec q} '}^{(+)} \rangle = \lim_{\epsilon \to 0} i
\epsilon G ' (E_q + i \epsilon)
\, | \vec q \, \rangle \quad ,
\eeq
lie in the $\eta$ ($\lambda$)--space,
when the corresponding asymptotic momentum $\vec q$ belongs
to the $\eta$ ($\lambda$)--space,
respectively.
Here,  $ | \vec q \,\rangle $ is a momentum eigenstate and
$H_0 \, | \vec q \,\rangle = E_q \, | \vec q \,\rangle $.
Immediately, the question arises
what is   the connection between the scattering states
$ | \Psi_{\vec q}^{(+)} \rangle
\equiv \lim_{\epsilon \to 0} i \epsilon G (E_q + i \epsilon)
\, | \vec q \,\rangle $ and $| {\Psi_{\vec q} '}^{(+)} \rangle $?
It is not obvious that eq.~(\ref{conn}) holds also for these
scattering states, which are defined through  specific boundary conditions.
We sketch now a proof, showing
that eq.~(\ref{conn}) is indeed valid in this case and that
the relation
\beq
\label{prov}
| {\Psi_{\vec q} '}^{(+)} \rangle = U^\dagger | {\Psi_{\vec q} }^{(+)} \rangle
\eeq
is satisfied.
The following steps appear highly plausible but do not replace a mathematically
rigorous proof. Consider the left hand side of this equation:
\beqa
\label{15}
| {\Psi_{\vec q} '}^{(+)} \rangle &\equiv&
\lim_{\epsilon \to 0} i \epsilon G ' (E_q + i \epsilon)
\, | \vec q \rangle = \lim_{\epsilon \to 0} i \epsilon
U^\dagger \, G  (E_q + i \epsilon) \, U \, | \vec q \,\rangle  \nonumber \\
&=& U^\dagger \, \lim_{\epsilon \to 0} i \epsilon G  (E_q + i \epsilon)
\, \left( \Big( \eta + \lambda A \eta \Big)
\left( 1 + A^\dagger A \right)^{-1/2} \right.\\
&& {} +\left.
\left( \lambda - \eta A^\dagger \lambda \right)
\left( 1 + A A^\dagger \right)^{-1/2}
\right)  \, | \vec q\, \rangle~. \nonumber
\eeqa
In the last line we used eq.~(\ref{Uokubo}).
Let us define the operators $B$ and $C$ by
\beqa
\label{matrb}
B &\equiv & \eta B \eta = \left( \eta + A^\dagger A \right)^{-1/2} - \eta~, \\
C &\equiv & \lambda C \lambda = \left( \lambda + A A^\dagger \right)^{-1/2}
 - \lambda~.
\eeqa
Then eq.~(\ref{15}) takes the form
\beq
\label{16}
| {\Psi_{\vec q} '}^{(+)} \rangle =
U^\dagger \, \lim_{\epsilon \to 0} i \epsilon G  (E_q + i \epsilon)
\left(  | \vec q \,\rangle  + F \,  | \vec q \,\rangle \right)
\eeq
with
\beq
F=B + C + A ( \eta + B ) - A^\dagger ( \lambda + C )~.
\eeq
Now the operator $A$ and as a consequence $B$ and $C$ depend on the
interaction $V$ and are linear in $V$ in lowest order. This is
obvious from eq.(\ref{eqa}).
Further it can be excluded that the state $ F | \vec{q}\, \rangle $
is an exact  continuum eigenstate of $H$ to the energy $E_q$,
which would be required
to generate a pole $\propto 1 /(i \epsilon)$ in the application of
$G (E_q + i \epsilon )$.
Only then the factor $i \epsilon $ could be cancelled. As a consequence
one obtains
\beq\label{Feq0}
\lim_{\epsilon \to 0} i \epsilon G  (E_q + i \epsilon) \, F \,
| \vec q\, \rangle =0~,
\eeq
and thus eq.~(\ref{prov}) is indeed satisfied.
One may also use a perturbative argument and expand the full
resolvent operator
\beq
\lim_{\epsilon \to 0} i \epsilon G  (E_q + i \epsilon) \, F \,  | \vec q\,
 \rangle
= \lim_{\epsilon \to 0} i \epsilon \left( \sum_{i=0}^\infty
\left( G_0  (E_q + i \epsilon) V
\right)^i \right) G_0  (E_q + i \epsilon)  \, F \,  | \vec q\, \rangle \quad .
\eeq
Because of the above mentioned property of $F$ it is clear that
$\langle \vec{p}\, | F | \vec{q}\, \rangle $ does not contain a
term proportional to
$\delta ( \vec{p} - \vec{q}\, )$ and therefore one can not generate
a pole term
$1 / (i \epsilon )$ through the application of $G_0 ( E_q + i \epsilon )$ onto
$F | \vec{q}\, \rangle $. Consequently, each single term of the series
on the right hand side of this
equation equals to zero, which  leads again to (\ref{Feq0}).
We thus have shown  that the scattering states $\Psi_{\vec{q}}^{(+)}$ initiated
by momenta $\vec{q}\,$ from the low momentum $\eta$--space are transformed into
${\Psi_{\vec{q}} '}^{(+)}$, which obey eq.(\ref{Seta}).
The corresponding $\lambda$ components of ${\Psi_{\vec{q}} '}^{(+)}$
are strictly zero.

After considering the scattering states, we now turn our attention
to the bound states. Again, the obvious question is:
Where do the transformed bound states of $H$ reside?
If the cut--off $\Lambda$ is sufficiently large then $A$ goes to zero.
This is a simple consequence of the fact that $V$ is assumed to fall off
sufficiently fast for
high momenta. Consequently $\lambda H ' \lambda$ is approximately
equal to $\lambda H \lambda$
which contains only a small portion of $V$ and
thus can not support bound states at all.
The transformed bound states have therefore to be solutions  of
eq.(\ref{Seta}).
For the physically reasonable choices we used in section~3 this turns out to
be true. 
It is not known to us of which value of $\Lambda$ this argument breaks down.
In section 3 we shall use an iteration procedure to solve the nonlinear equation 
(\ref{eqa}). It is known from \cite{sw} that the iteration converges only if
the lowest eigenvalue of $\lambda H \lambda$ is greater than the largest 
eigenvalue of the effective $\eta$ space Hamiltonian $\eta H ' \eta$. 
If one chooses a sufficient small cut--off $\Lambda$, than we can expect that 
$\lambda H \lambda$ will provide a bound state, which of course is not the physical one.
In such a case the iteration method has to break down and the nonlinear equation (\ref{eqa})
has to be solved in another manner. It is conceivable that then the true physical 
transformed bound state resides in the effective $\lambda$--space.

One can argue just in the same way to see the validity of the equation
(\ref{conn}) for the states $| {\Psi_{\vec q} }^{(-)} \rangle \equiv
\lim_{\epsilon \to 0} i \epsilon G (E_q - i \epsilon)
\, | \vec q \,\rangle $ and
 $| {\Psi_{\vec q} '}^{(-)} \rangle \equiv
\lim_{\epsilon \to 0} i \epsilon G ' (E_q - i \epsilon)
\, | \vec q \,\rangle $.
Therefore, the $S$--matrices in the original and transformed
problem are the same:
\beq
S_{\vec q \vec{q\,}'} ' \equiv \langle  {\Psi_{\vec q} '}^{(-)} |
{\Psi_{\vec{q\,}'} '}^{(+)} \rangle =
\langle  {\Psi_{\vec q} }^{(-)} | U U^\dagger |
{\Psi_{\vec{q\,}'}}^{(+)} \rangle =
\langle  {\Psi_{\vec q} }^{(-)} |
{\Psi_{\vec{q\,}'}}^{(+)} \rangle = S_{\vec q \vec{q\,}'}~.
\eeq
As a consequence the on--shell T--matrix element evaluated by means of
the Lippman--Schwinger (LS) equation
\beq
\label{teff}
T ' = V ' + V ' G_0 T '\quad ,
\eeq
yields exactly the same matrix elements as gained via the LS equation to
the original problem
\beq\label{torig}
T = V + V G_0 T \quad .
\eeq
Note that in eq.(\ref{torig}) one integrates over the whole (infinite)
momentum range whereas in eq.(\ref{teff}) only momenta up to the cut--off
$\Lambda$ are involved.

An important  observation is that in the subspace of  momenta below
the cut--off most local operators are non--local.
For an arbitrary local operator $O (\vec p_1, \vec p_2 ) = O (\vec p_1) \;
 \delta^3 (\vec p_1 - \vec p_2 )$
one obtains
in the transformed space
\beq
\tilde{O} (\vec p_1, \vec p_2\ ) = \int d^3 {p}'  \, U^\dagger (\vec p_1 ,
{\vec p \,}' \,) \,
O ({\vec p \,} ' ) \, U({\vec p \,}' , p_2)~,
\eeq
which in general contains the usual $\delta$--function part but in addition
also
a strong non--local piece.
These non--localities, which are easy to handle, are nothing but the
trace of the high momentum components from the full space.
Note, however, that the free particle Hamiltonian $H_0$ and as a consequence
the momentum operator $\vec p$ of a particle remain unchanged.
Certainly, one could also
unitarily transform the operators $H_0$ and $\vec p$. This would not
yield any new aspects
since the original and the unitarily transformed problems would be
trivially identical, the unitary transformation only leads to
change of basis. We shall not consider this trivial  case any further.
Physical observables
such as the deuteron binding energy and phase shifts are identical in
the original and transformed problems as shown before.
We remark that for instance the average momentum in the deuteron will change in
the
transformed problem because  the momentum operator is not unitarily
transformed.
This is not a problem since it is  not  direct observable.
However, all current operators have to be unitarily transformed and this
guarantees the equivalence of all  observables.

To end this section, we discuss an alternative way of determining the
operator $A$, as exhibited in ref.\cite{SO}.
That method uses the knowledge of the scattering states to the
original Hamiltonian $H$.
Let us now look in some more detail at this formalism.
As it was already pointed out above, the connection between the scattering
states in the original and transformed problem is given by
\beqa
\left| \Psi_{\vec{q}}^{(+)} \right\rangle &=& U
\left| {\Psi_{\vec{q}} '}^{(+)} \right\rangle  \\
&=& \left( \Big( \eta + \lambda A \eta \Big)
\left( 1 + A^\dagger A \right)^{-1/2} \, \eta
+ \left( \lambda - \eta A^\dagger \lambda \right)
\left( 1 + A A^\dagger \right)^{-1/2} \, \lambda
\right)  \, \left| {\Psi_{\vec{q}} '}^{(+)} \right\rangle~.  \nonumber
\eeqa
For momenta $\vec{q}$ from the $\eta$--space this equation takes a simpler form
\beq
\label{17}
\left| \Psi_{\vec{q} \; \in \; \eta}^{(+)} \right\rangle
= \Big( \eta + \lambda A \eta \Big) \left( 1 + A^\dagger A \right)^{-1/2}
\, \eta
\left| {\Psi_{\vec{q} \; \in \; \eta} '}^{(+)} \right\rangle \quad,
\eeq
since the resolvent operator of the transformed Hamiltonian $H '$
is block--diagonal as already pointed out before, cf eq.~(\ref{resolv}).
Using the trivial relation $( \eta + \lambda A \eta )
( \eta + \lambda A \eta ) =
\eta + \lambda A \eta$ we conclude from the eq.(\ref{17}) for all
momenta $\vec{q} \; \in \; \eta$ that
\beq
\lambda \left| \Psi_{\vec{q} \; \in \; \eta}^{(+)} \right\rangle
= \lambda A \eta \left| \Psi_{\vec{q} \; \in \; \eta}^{(+)} \right\rangle \quad
.
\eeq
Projecting this equation on to the state $\langle \vec{p}\, |$,
$p > \Lambda$, and making use of the  relation
\beq
\label{relat}
\langle \vec{p}_1 \,| \Psi_{\vec{p}_2}^{(+)} \rangle
= \delta ( \vec{p}_1 - \vec{p}_2 )
+ \frac{T (\vec{p}_1, \vec{p}_2, E_{p_2} )}{E_{p_2} - E_{p_1}
+ i \epsilon} \quad ,
\eeq
where $ T(\vec{p}_1, \vec{p}_2, z )$ is  the usual off--shell
$T$--matrix,
we end up with the following linear integral equation for the operator $A$:
\beq
\label{linear}
A (\vec{p}, \vec{q} ) = \frac{T (\vec{p}, \vec{q}, E_q)}{E_q - E_p}
- \int  d^3 q '  \, \frac{ A ( \vec{p}, \vec{q\,}' )
\; T (\vec{q\,}', \vec{q}, E_q )}
{E_q - E_{q '} + i \epsilon} \quad .
\eeq
Here the integration over $q '$ goes from 0 to $\Lambda$.
Note this is not a usual
Lippmann--Schwinger equation, since the position of the pole, $E_q$,
in the integration
over $q '$ is not fixed but moves with $q$. It is the second argument
in $A$ which varies, the first one is a parameter for the integral equation.

\section{Applications}
\setcounter{equation}{0}

This section is split into various paragraphs. First, we discuss
in detail how to determine the operator $A$ for a given realistic
potential. We then compare the results obtained in the space of
momenta below the chosen cut--off with the ones in the unrestricted
Hilbert space. Since the projection formalism is exact, we can recover
all physical quantities obtained from the full potential to an arbitrary
accuracy in the smaller space with low momenta only.
This extends the results exhibited in the
letter~\cite{egmplb}.
Third, we use this model to study the expansion
of short--range physics in terms of contact terms and draw some
conclusions for the application of chiral effective field  theories.
All these calculations are naturally done in momentum space. For
illustration, we finally show some results in coordinate space.

\subsection{Basic model and determination of the operator $A$}

Our starting point is a model potential which captures
the essential features of the nucleon--nucleon (NN) interaction.
We choose a momentum space NN potential with an attractive and a repulsive
part corresponding to the exchange of a light and a heavy scalar
meson, their masses denoted by $\mu_L$ and $\mu_H$,
respectively
\begin{equation}
\label{PotMT}
V (\vec{q}_1, \vec{q}_2) = \frac{1}{2\pi^2} \biggl(
\frac{V_H}{t + \mu_H^2} -
\frac{V_L}{t + \mu_L^2} \biggr)~,
\eeq
with $t = (\vec{q}_1 - \vec{q}_2)^2$. The strengths of the meson
exchanges parametrized by  $V_{L}$ and $V_H$, respectively, will
be determined later.
This potential still contains all partial waves.
The numerical investigations will be restricted to NN S--waves.
One  can work out the corresponding S--wave potential
in closed form,
\beq
\label{swave}
V (q_1, q_2) = \frac{1}{2 \pi q_1 q_2}
\left(  V_H \; \ln \left( \frac{(q_1 + q_2)^2 + \mu_H^2}{(q_1 - q_2)^2
+ \mu_H^2} \right)
-  V_L \; \ln \left( \frac{(q_1 + q_2)^2 + \mu_L^2}{(q_1 - q_2)^2
+ \mu_L^2} \right) \right)~,
\eeq
with  $q_{1,2}= | \vec q_{1,2} |$.
The nonlinear equation (\ref{eqa}) simplifies correspondingly.
Still, it can only be solved  numerically. We do this by iteration
starting with
\beq
\label{eq:start}
A = \frac{V (\vec{p}, \vec{q}\,)}{E_{\vec{q}} - E_{\vec{p}}}~.
\eeq
After four iterations we perform
an average over the values of the operator $A$ with suitably chosen weight
factors. The choice of these weight factors is responsible for
the quick convergence of this iteration method.
A typical total number of iterations is 40 to achieve an accuracy of 0.0001 GeV$^{-3}$
for the function $A (p, q)$.
The function $(E_q -E_p)$ in eq.~(\ref{eqa}) multiplying $A (p, q)$
requires some caution when $p$ and $q$ go to $\Lambda$.
We proceed by regularizing  the original potential $V (k ',k)$. We multiply it
with some smooth functions $f(k ')$ and $f(k)$ which are zero in a narrow
neighborhood of the points $k '= \Lambda$ and $k = \Lambda$ and one elsewhere.
The precise form of this regularization does in fact not matter as already
argued in~\cite{egmplb}. However, for the actual calculations presented
here, we choose
\beqa
f (k) &=& 1 , \mbox{\hskip 5.5 true cm}  \mbox{for} \quad
 k \leq \Lambda - a\quad \mbox{and} \quad
k \geq \Lambda + a~, \nonumber \\
f ( k) &=& \frac{1}{2} \left( 1 + \cos \left( \frac{\pi (k -\Lambda + a)}{b}
\right) \right) ,
\quad \quad \mbox{for} \quad  \Lambda -a \leq k \leq \Lambda -a+b~, \\
f ( k) &=& \frac{1}{2} \left( 1 + \cos \left( \frac{\pi (k -\Lambda - a)}{b}
\right) \right) ,
\quad \quad \mbox{for} \quad  \Lambda +a -b \leq k \leq \Lambda +a~,
 \nonumber \\
f ( k) &=& 0 , \mbox{\hskip 5.5 true cm} \mbox{for} \quad \Lambda -a+b \leq k
\leq  \Lambda +a-b~, \nonumber
\eeqa
with $a$ and $b$  parameters of dimension [energy].
This modification of the potential $V$ is depicted in fig.~\ref{fig1}.
Now $A (p, q)$ based on that modified potential is well defined for $p, q \;
\rightarrow \Lambda$.
\begin{figure}[Htb]
\vspace{-0.4cm}
\centerline{\psfrag{zzz}{\hskip -1.6 true cm \small
$V^{\rm mod} (q, q ')$ [GeV$^{-2}$]}
\psfrag{zq1}{\hskip -0.3 true cm \small $q \;$ [GeV]}
\psfrag{zq2}{\small $q ' \;$ [GeV]}
\psfig{file=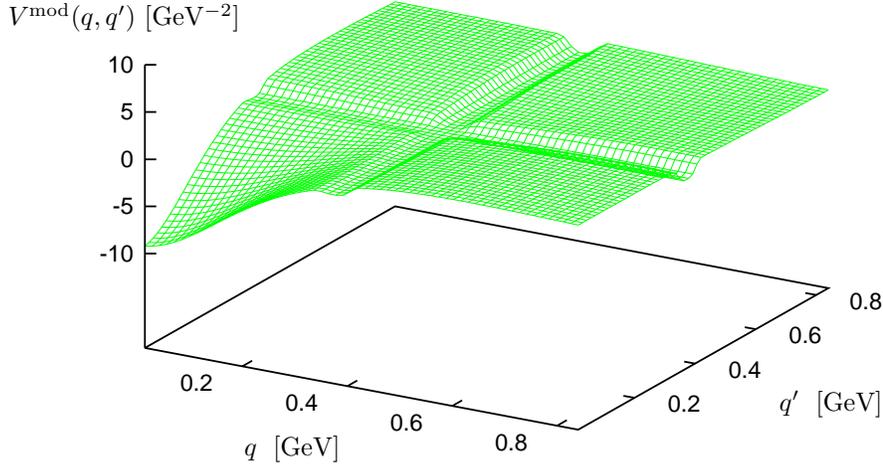,width=5in}}
\caption[potreg]{\label{fig1}Modification of the potential due to the
regularization.}
\end{figure}
To further quantify the effect of this regularization,
we set for simplicity  $b=0$.
Then it is easy to estimate the difference of  the
regularized on--shell $T$--matrix, $T^{\, \rm reg} (q, q)$,
to the unregularized (exact) one, $T (q,q)$, by
\beq
T^{\, \rm reg} (q, q) = T (q, q)
- a \frac{2 \Lambda^2 V (q, \Lambda) T (\Lambda, q)}
{E_q - E_\Lambda}  + {\cal O} ( a^2) ~,
\eeq
for all $q < \Lambda -a$ and $ q > \Lambda +a$. Similarly,
the modification of the deuteron binding energy can be obtained
by calculating the expectation value of $\langle \Psi_D | V -
V^{\, \rm reg} | \Psi_D \rangle$:
\beq
E^{\, \rm reg} = E- 4 a \Psi_D (\Lambda ) \Lambda^2  \int_0^\infty
\, k^2 \, d k \, V( \Lambda, k )  \Psi_D (k) + {\cal O} ( a^2) ~,
\eeq
where $\Psi_D (k)$ is the momentum space deuteron wave function.
Thus for any fixed $q< \Lambda -a$ and infinitesimal $\epsilon > 0$ one can
choose $a$ such that
$| T^{\, \rm reg} (q, q)  -  T (q, q) | < \epsilon$ and
$| E^{\, \rm reg} -  E | < \epsilon$.
Later we shall give numerical examples of the effects caused by
that potential modification in some
specific cases. We show later numerically that the effective potential $V ' (q,
q ')$
is affected only within the width $a$ for $q, q ' \; \rightarrow \Lambda$.
There both $V '$ and $V$ go to zero and have no effect on the observables.

We now have to fix the parameters for the potential $V$. We choose
these as given in ref.\cite{ETG}: $V_H = 7.291$, $V_L = 3.177$,
$\mu_H = 613.7$~MeV and $\mu_L = 305.9$~MeV.\footnote{For the light meson,
we could have chosen the pion mass.
However, since nuclear binding is largely due to correlated two--pion
exchange, a somewhat larger value was chosen. All conclusions drawn
in what follows are, however, invariant under the precise choice of
this number.}
That potential supports one bound state at $E= -2.23$ MeV and leads to
S--wave phase shifts in the  $^3S_1$ partial wave in fair agreement with
results
from NN partial wave analysis. Thus
this potential captures some essential features of the NN interaction.
Next, we have to select a value for the
cut--off $\Lambda$. In principle, $\Lambda$ could take any value,
in particular also above
the larger of the two effective meson masses in the potential. We shall
comment on that below. We are, however, mostly interested in an
effective theory with small momenta only and thus we start by setting
$\Lambda = 400$ MeV. To be more precise, by ``small'' we mean a scale
which is below the mass of the exchanged heavy particle so that one
can consider the situation with a propagating light meson and the
heavy meson integrated out and substituted by a string of local
contact terms (as will be discussed in more detail below).

The nonlinear equation is discretised using Gauss--Legendre quadrature
points. Here we have choosen $a=20$ keV, $b=10$ keV  and
100 Gauss--Legendre  points. The shift in the binding energy due to the
regularization is 0.012 KeV, which is a 0.01 permille effect.
Note that it can be made smaller if so desired. The resulting $A(p,q)$ is
shown in fig.~\ref{fig2}.
\begin{figure}[Htb]
\vspace{-0.8cm}
\centerline{\psfrag{zzz}{\hskip -1.4 true cm \small $A (p, q )$ [GeV${}^{-3}$]}
\psfrag{zq}{\small $q \;$ [GeV]}
\psfrag{zp}{\small $p \;$ [GeV]}
\psfig{file=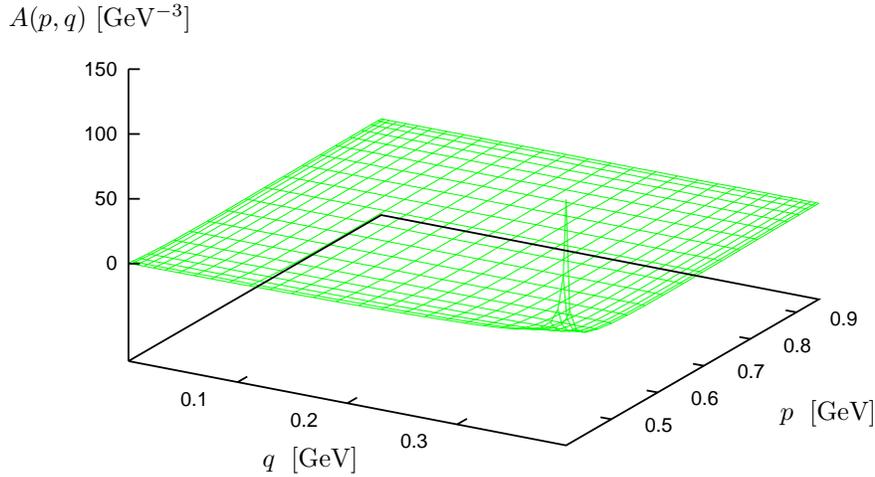,width=5in}}
\caption[matrixa]{\label{fig2}The function $A (p, q)$ for $\Lambda =400$ MeV.}
\end{figure}
Here we have  not shown the function $A(p,q)$ in the region of
regularization, i.e. for $p<\Lambda + a$, $q > \Lambda -a$. In fact
in this region the function $A (p,q)$ goes smoothly to zero. This is
guaranteed by the regularization  and can be seen explicitly from
eq.(\ref{eqa}).

\subsection{Effective potential, scattering and bound states}

Having calculated the operator $A$, we are now in the position to
consider observables. First, we need the transformed Hamiltonian $H'$.
The determination of $H '$ according to eq.(\ref{14})
requires the calculation of $( \eta+ A^\dagger A)^{-1/2}$.
This is done in the following manner: as already described
in  section~2,
we first define the function $B (q, q')$ as given in eq.(\ref{matrb}).
Consequently, this function  has  to satisfy the
following nonlinear equation as follows from  simple algebra:
\beqa
\label{equatb}
B (q, q') &=& -\frac{1}{2} \int_\Lambda^\infty \,  p^2 \, d  p \,
A (  p, q ) A (p, q ') - \frac{1}{2}
\int_0^\Lambda \, {\tilde q}^2 \, d \tilde q \,
B (\tilde q, q) B (\tilde q, q')  \\
&& {} - \int_\Lambda^\infty \,  p^2 \, d  p \,
\int_0^\Lambda {\tilde q}^2 \, d \tilde q \,
A (  p, q ) A (p, \tilde q ) \left( B (\tilde q, q')  +
\int_0^\Lambda {\tilde{q} ' \,}^{2} \, d {\tilde q} '\, B
(\tilde q, {\tilde q} ') B ({\tilde q} ', q')
\right) \quad . \nonumber
\eeqa
The function $B (q, q')$ defined in eq.(\ref{matrb}) is
obviously symmetric in the arguments $q$, $q '$:
$B (q, q') = B (q ', q)$. We have solved
equation (\ref{equatb}) by iteration, using the same Gauss--Legendre
quadrature
points as discussed before and starting with $B = -(1/2) \, A^\dagger A$.
We have found a very fast convergence of the iteration method in this case.
The integrations present in eq.(\ref{14}) to determine $H'$
are performed  by standard  Gauss--Legendre
quadratures and we end up with an effective potential $V '(q ',q)$ defined for
$q$, $q ' \leq\Lambda$. It is displayed in fig.~\ref{fig3}
in comparison to the original underlying potential for $\Lambda = 400\,$MeV.
\begin{figure}[htb]
\vspace{-0.4cm}
\centerline{\psfrag{zzz}{\hskip -1.4 true cm \small $V (q, q ')$
[GeV${}^{-2}$]}
\psfrag{zq1}{\small $q \;$ [GeV]}
\psfrag{zq2}{\small $q ' \;$ [GeV]}
\psfig{file=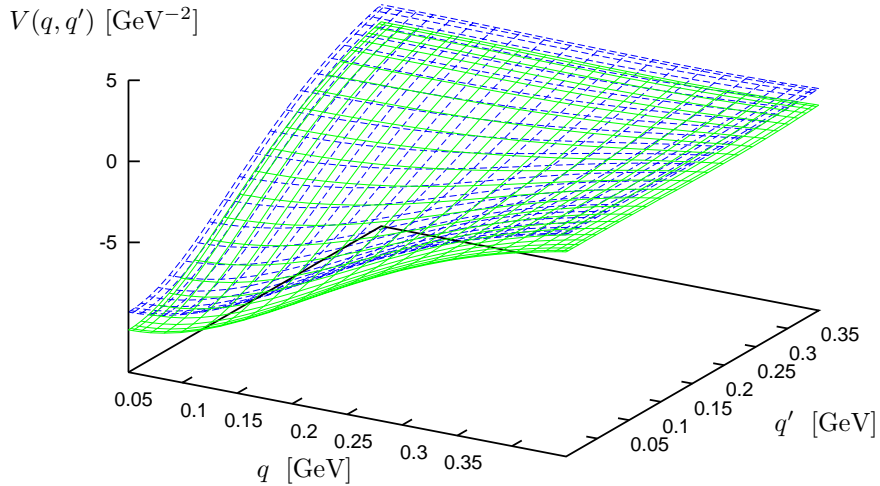,width=5in}}
\caption[potreg]{\label{fig3} Effective two--nucleon potential (solid
green lines) in
comparison with the original potential (dashed blue lines) for momenta less
than
400 MeV.}
\end{figure}
Note that the region of the regularization is again not shown to keep
the presentation  clearer.
One finds that the effective and the original potentials have a
similar shape for momenta below the cut--off.
The main effect of integrating out of high momentum components
at the level of the potential seems to be given in this case
just by  an overall shift.

\begin{figure}[htb]
\vspace{0.4cm}
\parbox{8cm}{
\centerline{\psfrag{zzz}{\small $\delta \; [ \, {}^\circ \,]$}
\psfrag{zen}{\small $T_{\rm lab} \;$ [MeV]}
\psfig{file=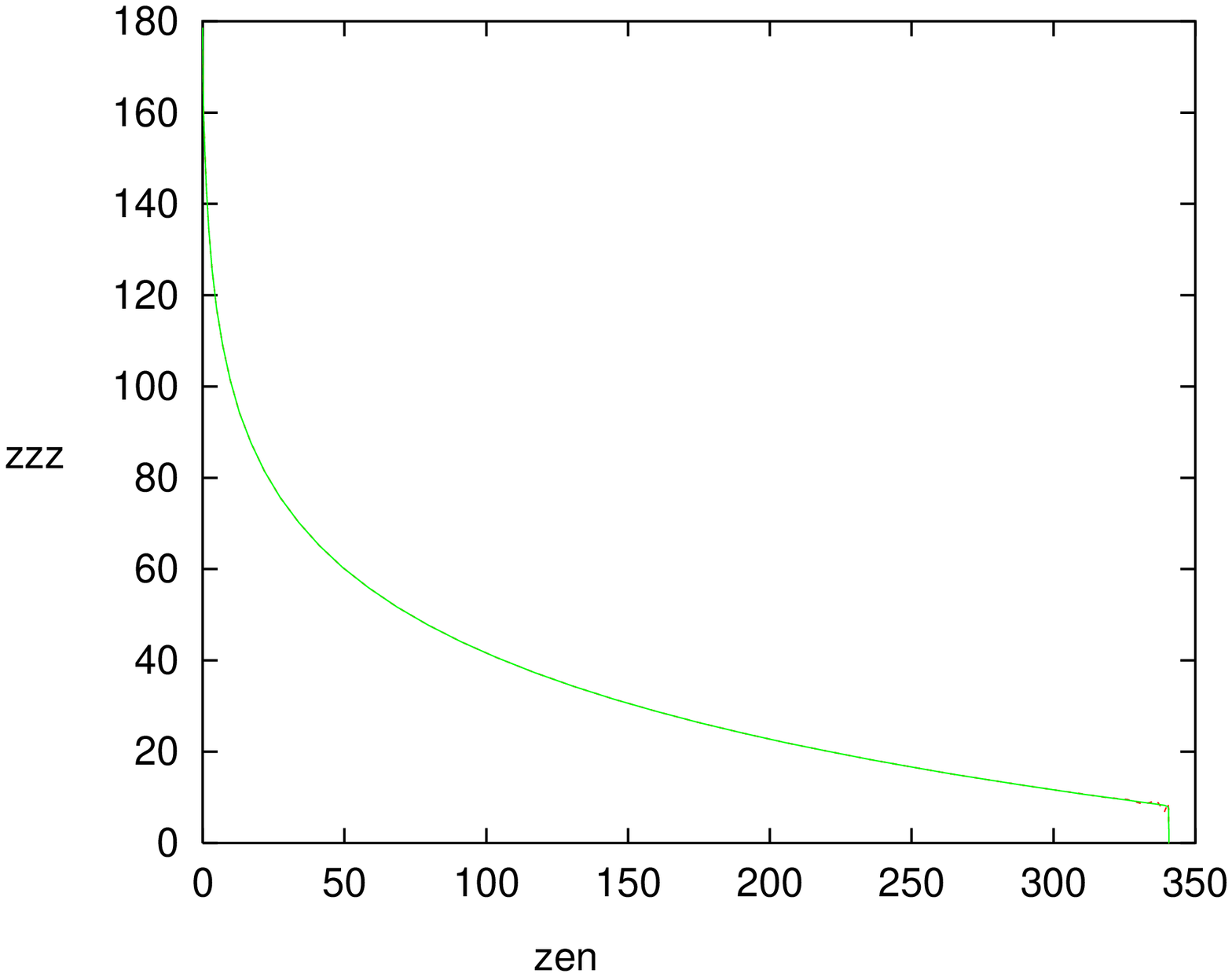,width=2.9in}}}
\hfill
\parbox{8cm}{
\centerline{\psfrag{zzz}{\small $\delta \; [ \, {}^\circ \,]$}
\psfrag{zen}{\small $T_{\rm lab} \;$ [MeV]}
\psfig{file=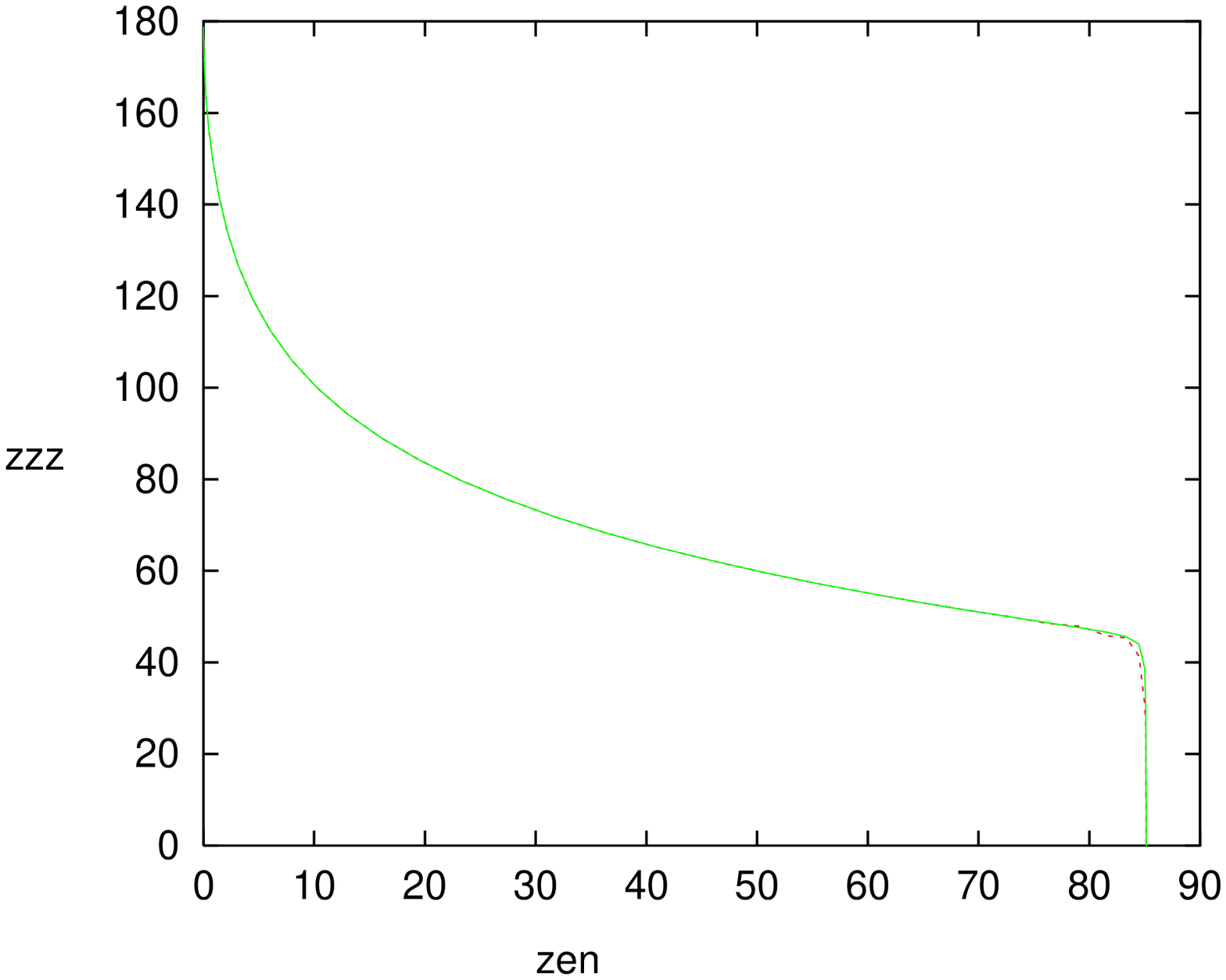,width=2.9in}}}
\vspace{0.51cm}
\caption[potreg]{\label{fig4} Phase shifts from the effective potential
(red dashed line) and
the original potential (green solid line) as a function of the kinetic energy
in the lab frame. Left (right) panel: $\Lambda =$400 (200) MeV.}
\end{figure}
The solution of the effective LS equation~(\ref{teff}) is now
very simple since the integration is confined to $q \leq \Lambda$.
The effective bound state wave function
obeys a corresponding homogeneous integral equation
\beq
\label{effschroed}
\frac{q^2}{ m_N} \Psi (q) + \int_0^\Lambda \, {\tilde q}^2 \, d \tilde q \,
V '(q, \tilde q ) \Psi (\tilde q) = E \Psi (q)~,
\eeq
with $m_N = 938.9\,$MeV the nucleon mass.
Using 40 quadrature points the resulting binding energy agrees within 9 digits
with the result gained from the corresponding homogeneous equation driven by
the original potential $V$ and defined in the whole momentum range.
Furthermore,  the S---wave phase
shifts agree perfectly  solving either eq.(\ref{torig}) in the
full momentum space or eq.(\ref{teff}) in the space of low momenta only.
This is shown in fig.~\ref{fig4}.
Note that due to the regularization the phase shifts go to zero for
$q \to \Lambda$ and that in the figure, the phase shifts are shown
as a function of the kinetic energy in the lab frame and the zero
occurs at $T_{\rm lab} = 2\Lambda^2/m_N = 341\,(85)\,$MeV for $\Lambda =
400\,(200)\,$MeV.

One can repeat this numerical exercise choosing other cut-off values.
Setting for instance $\Lambda = 2$ GeV,
the corresponding function $A(p,q)$ is shown in fig.~\ref{matrixa1}
and the effective potential $V '(q ',q)$ is rather close to the original one,
$[ V (q ', q) - V ' (q ', q)]/V(0,0) \sim 0.02$.
Here we have choosen
$a=$200 keV and $b=$100 keV, which leads to the same shift in the deuteron
binding
energy as in the case with $\Lambda =$400 MeV.
\begin{figure}[htb]
\vspace{-0.4cm}
\centerline{\psfrag{zzz}{\hskip -1.9 true cm \small $A (p, q )$
 [GeV${}^{-3}$]}
\psfrag{zq}{\small $q \;$ [GeV]}
\psfrag{zp}{\small $p \;$ [GeV]}
\psfig{file=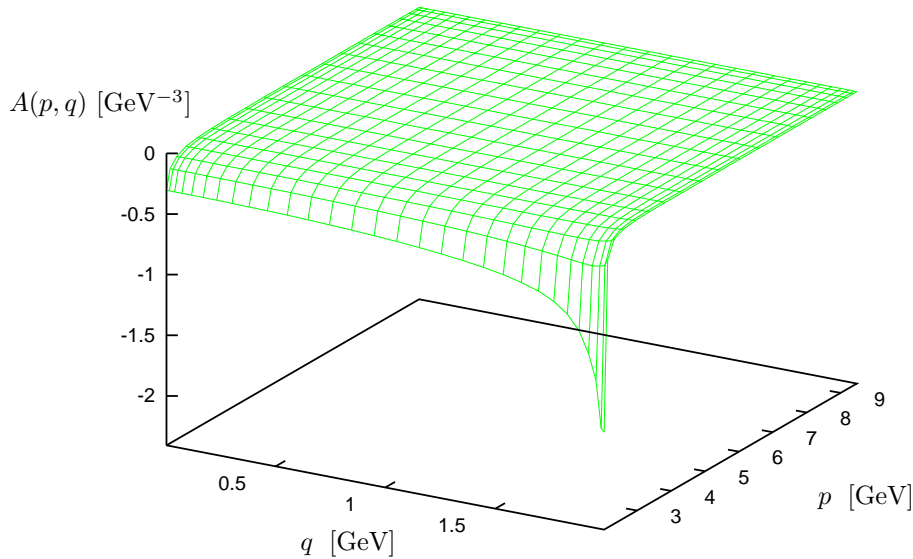,width=5in}}
\caption[matrixa1]{\label{matrixa1}The function $A (p, q)$
for $\Lambda=2$ GeV.}
\end{figure}
The closeness of the effective potential with the original one
for such a large cut--off value had to be expected, since the higher momenta
play only a perturbative role. On the other hand for rather small cut-off
values of $\Lambda$, like e.g. 200~MeV, the effective two-nucleon potential
is quite different from the original one. This is shown in fig.~\ref{fig5}.
\begin{figure}[htb]
\vspace{-0.4cm}
\centerline{\psfrag{zzz}{\hskip -1.4 true cm \small $V (q, q ')$
[GeV${}^{-2}$]}
\psfrag{zq1}{\hskip -0.3 true cm \small $q \;$ [GeV]}
\psfrag{zq2}{\small $q ' \;$ [GeV]}
\psfig{file=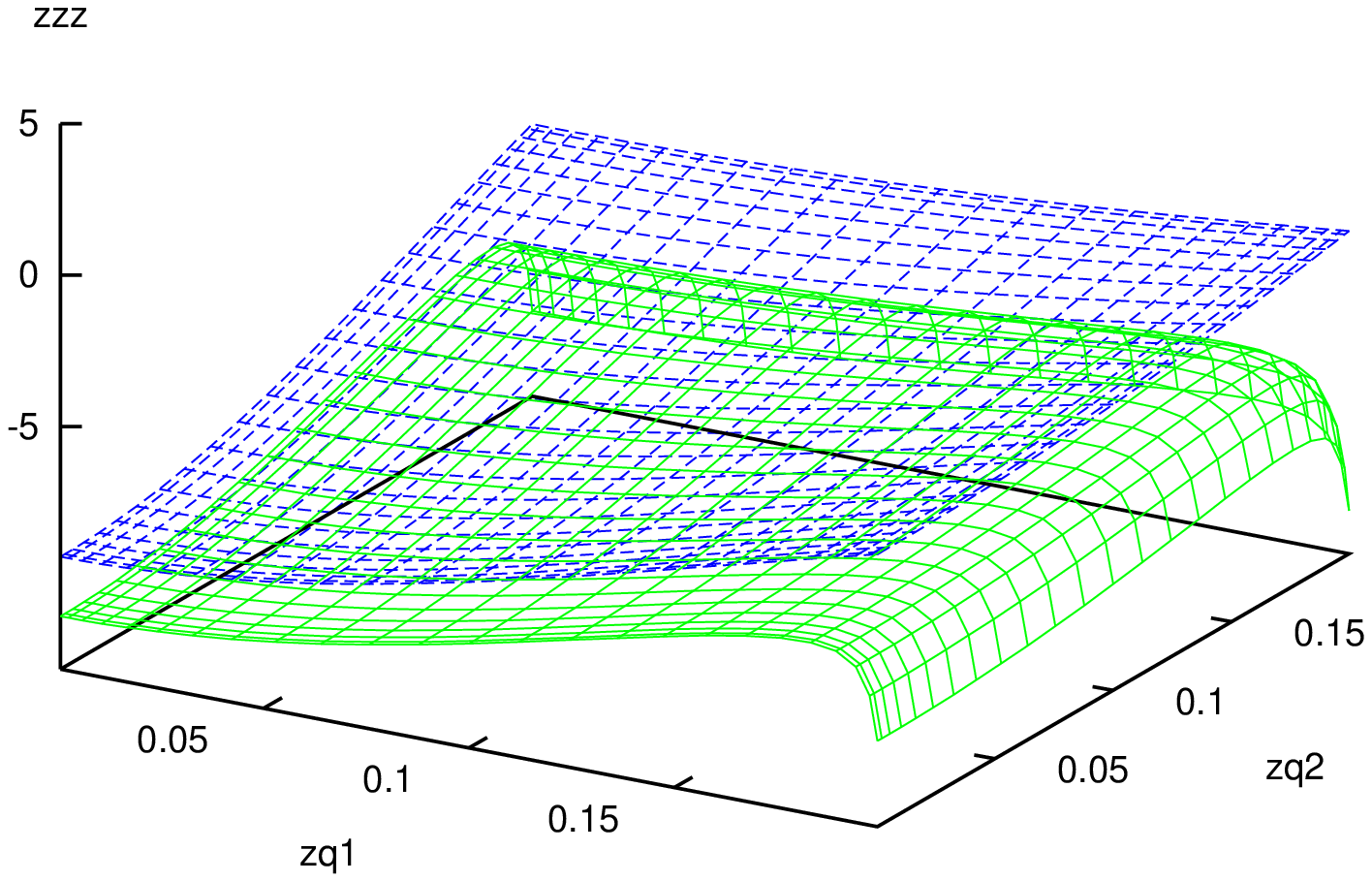,width=5in}}
\caption[potreg]{\label{fig5} Effective two--nucleon potential (solid 
green lines) in
comparison with the original potential (dashed blue lines) for momenta less
than 200 MeV.}
\end{figure}
Again for $\Lambda = 200$ MeV the binding energy and the phase shifts
agree perfectly using the effective and the original formulation,
see fig.~(\ref{fig4}). We have found solutions up to 
$\Lambda \simeq 100\,$MeV.  For smaller
cut--off values, the effective potential develops so much structure
that one would have to modify the method of solving the integral
equation to determine $A$. For the reasons mentioned above, we do not
pursue this issue here any further. All that demonstrates
that the nonlinear decoupling equation can be solved reliably and
consequently, the effective Hamiltonian including all the physics
can be determined to arbitrary high precision.

To the best of our knowledge this is the first time that this projection
formalism has been applied to such a case. In the context of the nuclear
many--body problem, one  often uses a formalism which decouples two spaces,
one defining a model space of low--lying states and the other the rest of the
$N$--particle Hilbert space (the so--called $P$ and $Q$ spaces).
The aim is to derive an effective interaction
acting in the model space only. This should lead to the exact
spectrum of the low--lying states. To do that a suitable
transformation is performed to decouple the
two spaces. It leads to a decoupling equation of exactly the form given
in eq.(\ref{eqa}),
however, acting now  in the space of $N$ particles. This problem is
very hard to solve and approximations are needed. In approximate schemes
that nonlinear equation is often reformulated into a linear form on a
two-body level using the exactly known interacting two-body states
(some references are~\cite{kuo}\cite{SO}). This is also
a feasible way to proceed in our context as was displayed at the end
of section~2.
Because we consider  two particles only, it is an exact reformulation
and the operator
$A$ obeys the linear equation (\ref{linear}).
We have solved that equation for an S--wave. For that,
the function $A (p, q)$ was expanded into cubic splines $S_k (q)$ \cite{spli}
\beqa
\label{splines}
A (p, q) = \sum_k S_k (q) \, A (p, q_k)
\eeqa
based on a set of suitably chosen gridpoints $\{ q_k \}$ in
the interval $[ 0, \, \Lambda ]$. This procedure leads to a system of linear
equations
for the expansion coefficients $A (p, q_k)$, which is solved by standard
methods. Consequently, the dynamical input is the S--wave t--matrix $T (q', q,
E_q)$.
The resulting $A (p, q)$ agrees perfectly  with the one obtained
by solving the nonlinear equation.

\subsection{Implications for chiral effective field  theories}

Now let us establish some contact to the theory of nuclear forces in chiral
perturbation theory.
In this approach only the nucleons and lightest mesons (i.e. pions)
are typically kept
in the theory as explicit degrees of freedom.
The structure of the pion--nucleon  interaction
is constrained by the spontaneously broken chiral symmetry.
To take into account the effects of heavier degrees
of freedom such as heavier mesons and baryon resonances, contact
interactions between the nucleons are introduced in the corresponding
Lagrangian. Such contact interactions are not constrained by  chiral symmetry.
Starting from the most general Lagrangian for pions and nucleons and
using   power--counting rules one can
obtain  formal expressions for the effective nucleon--nucleon potential
(see for instance~\cite{bira}\cite{EGM}),
which has to be put into the Lippmann--Schwinger
equation to generate the corresponding S--matrix.
Note that  regularization and
renormalization are  needed to treat the ultraviolet divergent
integrals in the Lippmann--Schwinger  equation.
Furthermore, the pionic loops in the effective NN--potential should be
regularized consistently with the regularization of the
Lippmann--Schwinger equation.
This regularization is necessarily nonperturbative:
the presence of  a low--energy bound state signals the failure of  perturbation
theory. Once the theory is regularized (for instance with some cut--off)
all parameters in the Lagrangian may be fixed by fitting to data, like
e.g. the low energy NN partial waves and deuteron properties.
It is commonly believed that the coupling constants
scaled by some effective mass parameter $\Lambda_{\rm scale}$
should have the property of "naturalness", which means they should be of the
order of one.  Only then this expansion makes sense and the
power--counting is self--consistent.
The value of the scale  $\Lambda_{\rm scale}$ is obviously closely
connected to the radius of convergence of this expansion.
Let us first consider  the simpler case when
the pionic degrees of freedom  are integrated out. In that case,
the low--energy NN interaction can be described entirely in terms of
contact terms  and one expects the scale $\Lambda_{\rm scale}$ to be
of the order of the pion mass $m_\pi$.
Therefore, all physical parameters which describe the NN phase shifts up
to center of mass momenta of the order $m_\pi$, such as the
scattering length $a$ and the effective range $r_e$, are expected to scale
like appropriate inverse powers of $m_\pi$.
This is, however, not the case in nuclear physics:
the NN scattering length in the ${}^1S_0$--channel takes an unnatural
large value, $a=(-23.714 \pm 0.013) \; {\rm fm} \gg 1/m_\pi$.
As a consequence, the expansion in powers of small momenta breaks down
at  momenta much below the scale $m_\pi$.
The  physics of this phenomenon is well understood (there is a virtual
bound state very near zero energy) and amounts
to a  fine tuning between  different terms when the corresponding phase
shifts are  calculated~\cite{bira2}\cite{ksw}.
To achieve such cancellations in the effective field theory calculations
one has to "fine tune" the parameters.
We shall see below how this "fine tuning" works in our model.

The situation is more complicated in the effective theory with pions
since different scales appear explicitely. That is why it is not clear
a priori what scale enters the coefficients in the Lagrangian.
Using a modified dimensional regularization scheme and  renormalization group
equation
arguments, the authors of~\cite{ksw} have argued that $\Lambda_{\rm scale}
\sim 300\,$MeV. On the other hand it was argued by the
Maryland group~\cite{silas} that no useful and systematic
effective field theory (EFT) exists for two nucleons with
a finite cut--off as a regulator.
A systematic EFT is to be understood in the sense that the contributions
of the higher--order terms in the effective Lagrangian to the observables
at low momenta (i.e. the quantum averages of such operators) are small and
therefore truncation of such operators at some finite order is justified.
Within our realistic model, we can address this question in a quantitative
manner as shown below.

Let us now apply the concept of the effective theory to
our case.  The original potential plays the role of
the underlying theory of NN interactions in analogy to QCD underlying the
true EFT of NN scattering. Integrating out the momenta above
some cut--off $\Lambda < \mu_H$, we arrive via the unitary
transformation at the effective potential $V' (q,q')$. In analogy to
the true EFT we decompose $V'$ into two parts:
\beq\label{Vprime}
V' (q',q) = V_{\rm light}' (q',q) + V_{\rm contact}' (q',q)~.
\eeq
Here, $V_{\rm contact}'$ is a string of local contact terms of
increasing dimension, which is caused by the heavy mass particle and
the high momenta $p>\Lambda$. The piece $V_{\rm light}'$ is related to  the
light meson exchange, but again it is modified by integrating out the
high momenta and in general it will depend on the order to which we
expand $V_{\rm contact}'$. However, since we work in a model which
serves as an illustration, we simplify the procedure and keep $V_{\rm light}'$
fixed as the light meson exchange: 
$V_{\rm light} ' = V_{\rm light}$, where 
$V_{\rm light}$ denotes the second term in eq.~(\ref{swave}). 
Specifically,
\beq
\label{potcont}
V_{\rm contact}=V^{(0)}+V^{(2)}+V^{(4)}+V^{(6)}+\ldots~,
\eeq
with
\beqa
\label{const}
V^{(0)}&=&C_0~, \nonumber \\
V^{(2)}&=&C_2 ({q '}^2 + q^2)~, \nonumber \\
V^{(4)}&=&C_4 ({q '}^2 + q^2)^2 + C_4 ' {q '}^2 q^2~, \\
V^{(6)}&=&C_6 ({q '}^2 + q^2)^3 + C_6 ' ({q '}^2 + q^2) {q '}^2 q^2~,
\nonumber
\eeqa
and the superscript $'(2n)'$ gives the chiral dimension (i.e. the number
of derivatives).
Thus the first term on the right hand side of eq.(\ref{Vprime}) is the
purely attractive part
of the original potential due to the light meson exchange.
In this way,  we have an effective theory for
NN interactions, in which the effects of the heavy meson exchange
and the high momentum components
are approximated by the  series of  NN contact interactions
and the light mesons are treated explicitely.
Note that the constants $C_i$, $C_i '$ in eq.~(\ref{const}) correspond
to renormalized
quantities since the effective potential $V ' (q ', q)$ is regularized
with the sharp cut--off $\Lambda$.
The first question we address in our model is: what is  the
value of the scale $\Lambda_{\rm scale}$ and its relation to
the cut-off $\Lambda$? Of course, a priori these two scales are not
related. For the kind of questions we will address in the following,
we can, however, derive some lose relation between these two scales
as discussed below.
Since we  know $V '(q ',q)$ numerically, we can determine the constants $C_i$
by fitting the eq.(\ref{potcont}) to $V '(q ',q)-V_{\rm light} (q ', q)$.
This is done numerically using the standard FORTRAN subroutines for
polynomial fits to functions of one variable and taking the
corresponding polynoms
typically of order ten to eleven. Once this is done for the functions
$V(q, 0)$ and $V(q,q)$ all constants $C_i$, $C_i '$ in eq.~(\ref{const})
can be easily evaluated.
For the potential parameters used so far and the choices of $\Lambda=400\,$MeV
and  $\Lambda=300\,$MeV, the resulting constants are given in Table~1:

\begin{table}[h]
\centerline{\begin{tabular}{|c||c|c|c|c|c|c|} \hline
& $C_0$ & $C_2$ & $C_4$ & $C_4 '$ & $C_6$ & $C_6 '$  \\ \hline \hline
$\Lambda = 400$ MeV & 11.23 & -32.27 & 86.73 & 113.9 & -232.6 & -913.6 \\
\hline
$\Lambda = 300$ MeV & 11.15 & -32.93 & 85.76 & 115.6 & -265.2 & -783.7 \\
\hline
\end{tabular}}
\caption{The values of coupling constants $C_i$, $C_i '$
in [GeV${}^{(-2-i)}$] for
two choices of the cut--off  $\Lambda$.\label{tab1}}
\end{table}

\noindent
Naturalness of the coupling constants $C_i, C_i'$ means that
\beq
\frac{C_{2n}}{C_{2n+2}} =
a_n \Lambda_{\rm scale}^{2}~,
\eeq
where the $a_n$ are numbers of order one. Indeed, as one can read off
from table~\ref{tab1}, such a common scale exists, namely
\beq
\Lambda_{\rm scale} = 600~{\rm MeV}~.
\eeq
This  is a reasonable  value in the sense
that it is very close to the mass $\mu_H$ of the heavy mesons, which is
integrated out from the theory. Stated differently, the value
for $\Lambda_{\rm scale}$  agrees with naive expectations.
In fact, in the full theory (i.e. without the projection into the space
of small momenta only), this observation would be trivial since the expansion
of the heavy meson propagator gives
\beq
\frac{1}{t+\mu_H^2} = \frac{1}{\mu_H^2}\, \biggl( 1 - \frac{t}{\mu_H^2}
+ \frac{t^2}{\mu_H^4} +  \ldots \biggr) ~,
\eeq
\begin{figure}[htb]
\vspace{-0.4cm}
\centerline{\psfrag{zzz}{\hskip -1.6 true cm \small $V (q, q ')$ [GeV$^{-2}$]}
\psfrag{zq1}{\hskip -0.3 true cm \small  $q \;$ [GeV]}
\psfrag{zq2}{\small $q ' \;$ [GeV]}
\psfig{file=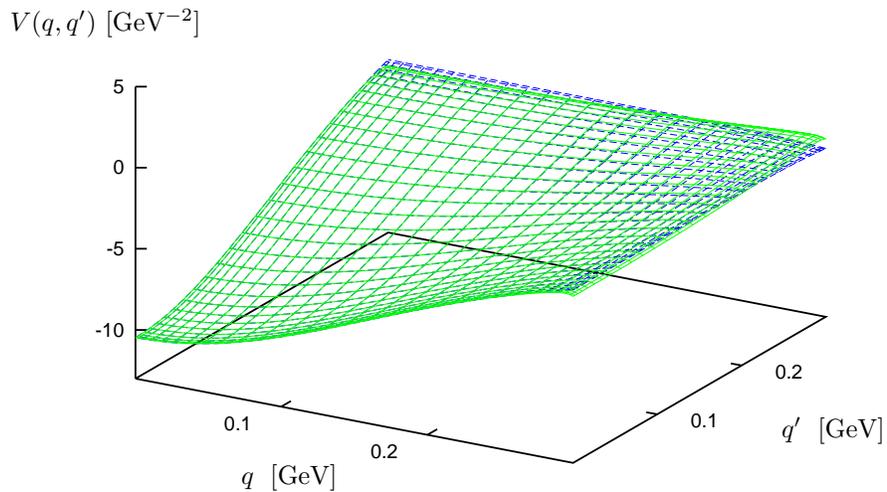,width=5in}}
\caption[potcont]{\label{fig6} Effective two--nucleon
potential $V ' (q ', q)$ (solid green lines) in
comparison with the truncated expansion (\ref{potcont})
(dashed blue lines) for $\Lambda=300$ MeV.}
\end{figure}
\noindent
and thus $\Lambda_{\rm scale} = \mu_H$. In our approach, however, the
large momentum components are mapped into the Hilbert space of small
momenta and thus a priori it is not obvious that $\Lambda_{\rm scale}$
is indeed given by $\mu_H$. As it turns out, even for cut--off values
like $\Lambda = 300\,$MeV, there is only a small difference between
the heavy meson mass and the ensuing natural mass scale.
Another observation is that the values of the $C_i$, $C_i '$ depend very
weakly on  the concrete choice of the cut--off $\Lambda$. Clearly, for such
an expansion of the heavy mass exchange in terms of contact interactions
 to make sense, $\Lambda$ has to be
chosen below $\mu_H$. Furthermore, since we explicitely keep the light
meson, $\Lambda$ should not be smaller than the mass $\mu_L$. If one were
to select such a value for the cut--off, one could also consider
the possibility of expanding the light meson exchange
in a string of contact terms. We do, however,
not pursue this option in here. Therefore, we conclude that one should
set $\Lambda < \Lambda_{\rm scale}$ but it is not possible to find a
more precise relation.
In fig.~\ref{fig6} we show how well the potential $V ' (q', q)$
is reproduced by the truncated expansion eq.(\ref{potcont})
for $\Lambda =300\,$MeV.

We have also calculated the two--body binding energy and the phase shifts using
the
form eq.(\ref{potcont}) with the constants given in table~\ref{tab1}.
The corresponding results are shown in fig.~\ref{fig7} and in
table~\ref{tab2}. The agreement with the exact
values is good. However, one sees that terms of  rather high order
should be kept in the effective potential $V_{\rm contact}$ in order to
have the binding energy correct within a few percent. Note, however,
that the value of the binding energy is unnaturally small on a typical
hadronic scale like the pion mass or the scale of chiral symmetry breaking.
The phase shifts are described fairly well for kinetic energies (in
the lab)  up to about 100~MeV
if one retains the first three terms in the expansion
eq.(\ref{potcont}). For $\Lambda = 400\,$MeV, one can not expect
any reasonably fast convergence for the binding energy any more. This can be
traced back
to the fact that one is close to the radius of convergence for momenta
close to the cut--off. More specifically, with $q = q' =400\,$MeV, the
pertinent expansion parameter is $({q'}^2 + q^2)/\Lambda_{\rm scale}^2
\simeq 0.9$. The ensuing very slow convergence is exhibited in
table~\ref{tab2}.
\begin{table}[h]
\centerline{\begin{tabular}{|c||c|c|c|c|c|} \hline
& $V^{(0)}$ & $V^{(0)}+V^{(2)}$ & $V^{(0)}+V^{(2)}+V^{(4)}$
& $V^{(0)}+V^{(2)}+V^{(4)}+V^{(6)}$ &
$V_{\rm contact}$   \\ \hline \hline
E [MeV] & 0.46 & 3.18 & 1.95 & 2.29 & 2.23 \\ \hline
E [MeV] & 0.67 & 7.15 & 1.82 & 3.15 & 2.23 \\ \hline
\end{tabular}}
\caption{The values of the binding energy calculated with $V_{\rm contact}$,
eq.~(\ref{potcont}),
for $\Lambda = 300\,$MeV  (second row)  and $\Lambda = 400\,$MeV (third row).
\label{tab2}\smallskip}
\end{table}
\begin{figure}[htb]
\vspace{0.4cm}
\parbox{8cm}{
\centerline{\psfrag{zzz}{\small $\delta \; [ \, {}^\circ \,]$}
\psfrag{zen}{\small $T_{\rm lab} \;$ [MeV]}
\psfig{file=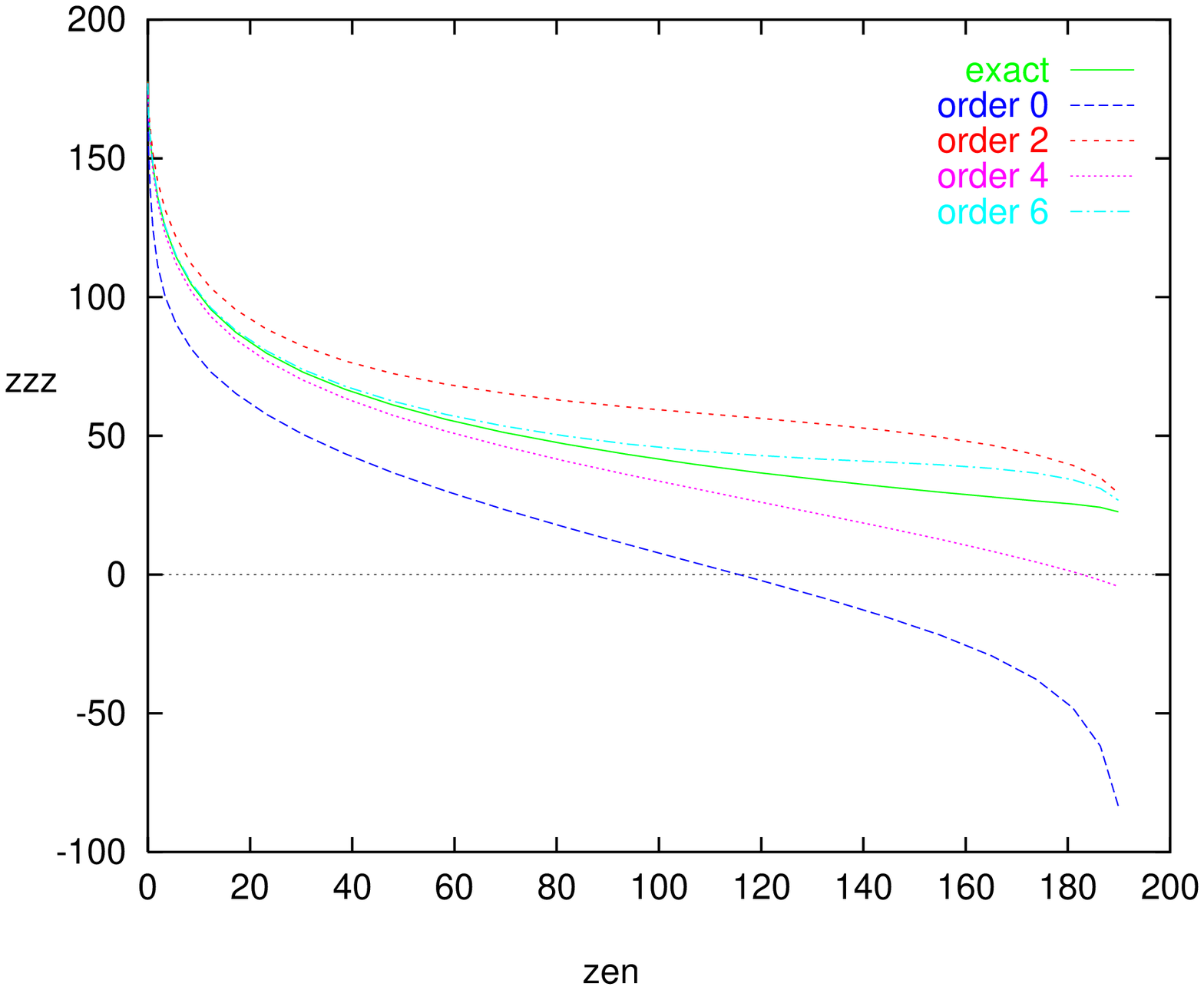,width=2.9in}}}
\hfill
\parbox{8cm}{
\centerline{\psfrag{zzz}{\small $\delta \; [ \, {}^\circ \,]$}
\psfrag{zen}{\small $T_{\rm lab} \;$ [MeV]}
\psfig{file=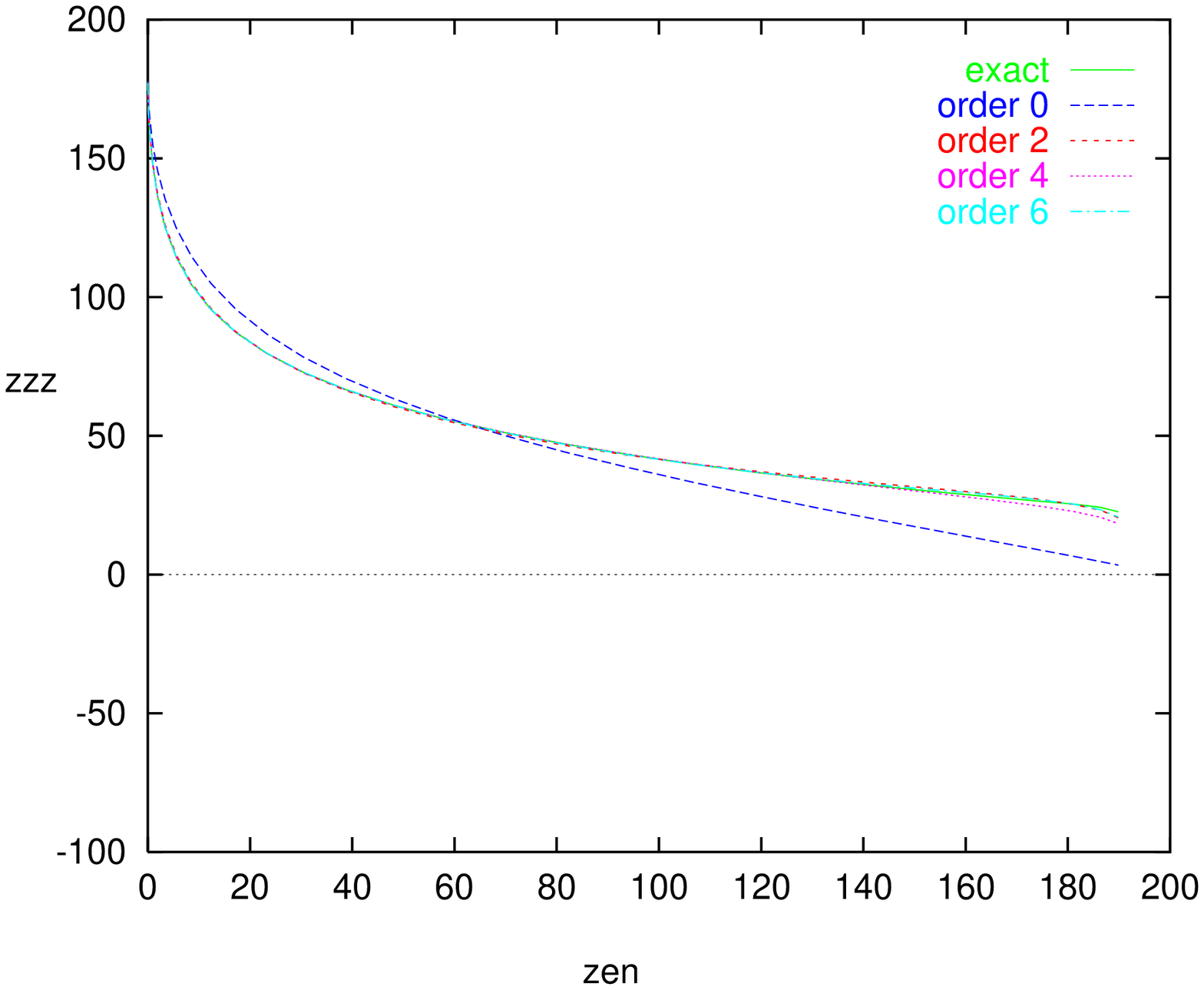,width=2.9in}}}
\vspace{0.51cm}
\caption[phcont]{\label{fig7} Phase shifts from the
effective potential $V '$ (solid line) and
the truncated expansion (\ref{potcont}) as a function of the kinetic energy
in the lab frame for  $\Lambda =$300 MeV. Left (right) panel: the
constants $C_i$,
$C_i '$ are fitted to the effective potential (to the NN phase shift).}
\end{figure}

Although we have found that $\Lambda_{\rm scale} \sim$ 600 MeV and
therefore the expansion of $V '$ in terms of contact terms seems to
converge for the chosen value of the cut--off $\Lambda=300$ MeV,
the ultimative test of convergence of the expansion (\ref{potcont})
would be to calculate the quantum averages of operators $V^{(0)}$,
$V^{(2)}$, $V^{(4)}$, $V^{(6)}$, $\ldots$, as it was proposed in~\cite{silas}.
This allows to directly draw conclusions about the size of such operators
in the effective action. 
In fact, the potential is not an observable and
naturalness of its coefficients can become meaningless when one itearates the
potential in the LS equation since in general large momenta are involved.
In our case, this can not happen and in the worst case, the convergence 
for observables will be controlled by the parameters $\Lambda/
\Lambda_{\rm scale}$ multiplied by coefficients of order one (with the
exception of quantites which require strong fine tuning).
These remarks will be substantiated by the results to be presented below.
Because of technical difficulties, the
authors of ref.\cite{silas}
have used for this check  the bound--state wave function in the ${}^1
S_0$--channel,
obtained from the zeroth--order potential and without pions
(in real world there is, of course, no bound state in the ${}^1 S_0$--channel).
This allows  only for a very rough estimate of the size of the operators
in eq.~(\ref{potcont}).
In our model we can perform numerically exact calculations of
these quantities using not only the bound--state wave function
but also the scattering wave functions.  Using the
relation eq.(\ref{relat}) one obtains for an arbitrary operator $O$
\beqa
\label{matrel}
\langle {\Psi_q '}^{(+)} | O | {\Psi_q '}^{(+)} \rangle = \langle q | O | q
\rangle
+ I_1 + I_1^* + I_2~,
\eeqa
with
\begin{eqnarray}
I_1 &=& \int \, {q '}^{2} \, d q '\,  O (q, q ') \,
\frac{T ' (q ', q, E_q)}{E_q - E_{q '} + i \epsilon}~, \nonumber \\
I_2 &=& \int \, {q_1 '}^{2} \, d q_1 ' \, {q_2 '}^{2} \, d q_2 ' \,
\frac{{T '}^* (q_1 ', q, E_q)}{E_q - E_{q_1 '} - i \epsilon} \,O(q_1 ', q_2 ' )
\,
\frac{{T '} (q_2 ', q, E_q)}{E_q - E_{q_2 '} + i \epsilon} \quad .
\end{eqnarray}
The results for quantum averages of the operators $V^{(2n)}$, $(n=0,1,2,3)$,
are  shown in  table~\ref{tab3}.
One observes a good convergence in agreement with the naive  expectation.
Indeed, since the cut--off is choosen to be $\Lambda =$300 MeV and the
scale $\Lambda_{\rm scale} \sim$600 MeV, one  expects the expansion
parameter to be of the order $\Lambda / \Lambda_{\rm scale} \simeq 1/2$.
Such a  value agrees well with the
one extracted from the results shown in
table~\ref{tab3}.
Note further that for higher energies, the convergence is slower,
which is also rather natural.
\begin{table}[h]
\centerline{\begin{tabular}{|c||c|c|c|c|} \hline
&$\langle \Psi | V^{(0)} | \Psi \rangle $ &
$\langle \Psi | V^{(2)} | \Psi \rangle $ &
$\langle \Psi | V^{(4)} | \Psi \rangle $ &
$\langle \Psi | V^{(6)} | \Psi \rangle $ \\ \hline \hline
deuteron &  25.94 MeV & -4.23 MeV & 1.07 MeV & -0.38 MeV \\ \hline
$E_{\rm lab} = 10$ MeV & 31.34 GeV${}^{-2}$ & -6.19 GeV${}^{-2}$ &
1.64 GeV${}^{-2}$ & -0.58 GeV${}^{-2}$ \\ \hline
$E_{\rm lab} = 50$ MeV & 11.54 GeV${}^{-2}$ & -3.28 GeV${}^{-2}$ &
1.11 GeV${}^{-2}$ & -0.44 GeV${}^{-2}$ \\ \hline
$E_{\rm lab} = 100$ MeV & 7.79 GeV${}^{-2}$ & -3.09 GeV${}^{-2}$ & 1.40
GeV${}^{-2}$ &
-0.71 GeV${}^{-2}$ \\ \hline
\end{tabular}}
\caption{The quantum averages of the operators  $V^{(0)}$,
$V^{(2)}$, $V^{(4)}$ and $V^{(6)}$ for the bound 
(second row) and
the scatterng states (third to fifth rows) for $\Lambda =300$ MeV.}
\label{tab3}
\end{table}

In the real world to which one applies the effective field theory,
one does not know the  true effective potential $V'$ and therefore also
the constants $C_i, C_i'$ are unknown.
They have to be determined by fitting such a type of effective potential to
the NN data, like the low energy phase shifts and/or the deuteron
binding energy. We can perform this exercise also in our case. Keeping again
$V_{\rm light} ( q ',q)$ as a seperate piece we have determined by trial and
error the
constants $C_0$, $C_2, \ldots$  by solving the homogeneous and
inhomogeneous LS equations  and truncating the series at various
orders.  The results can be summarized as follows: For $\Lambda =
300\,$MeV we restrict the fit to the phase shifts to momenta below
260~MeV. In that way, we avoid the distorsion of the phase shifts at
the edge of its kinematically allowed region in the model space. A fit
including only the first four terms, i.e. the constants $C_0, C_2,
C_4$ and $C_4'$ gives an excellent reproduction of the phase shifts as
shown in the right panel of fig.~\ref{fig7}. The corresponding
constants are $C_0 = 10.74$, $C_2 = -23.96$, $C_4 = 103.7$ and $C_4' =
-288.4$, all (with the exception of $C_4'$) within 25\% of the exact
values. The deviation of the fitted value for $C_4'$ from the exact
one reminds us of the fact that such fine tuning can produce sizeable
uncertainties in higher orders in such type of cut--off schemes and
thus the interpretation of such values has to be taken with some caution.
The bound state energy, which is not fitted, turns out to be
2.27~MeV, just 2\% above the exact value. If one includes the sixth
order terms in the fit, the results for the fourth and sixth order
coefficients become unstable. That can be traced back to the very
small contribution of these terms to the phase shift as long as $q <
260\,$MeV. A stable fit can be obtained by demanding that the
coefficient obey naturalness within a certain range, say within 20\%
or so. Clearly, to use polynoms of such high order to fit the smooth
phase shifts which are already very well approximated by the first four
terms does not make much sense. It is interesting to observe that the
fit itself tends to limit the terms in the chiral expansion giving a
good description of the phase shifts and binding energy. This
behaviour is reminiscent of the one found in the
dispersion--theoretical analysis of the nucleon form factors, where
one gets a best fit to the hadronic spectral function with a limited
number of vector meson poles (for details, see e.g.~\cite{hoeh}\cite{MMD}).

\begin{figure}[htb]
\vspace{-0.4cm}
\centerline{\psfrag{zzz}{\hskip -1.4 true cm \small $V (q, q ')$ [GeV${}^{-2}$]}
\psfrag{zq1}{\hskip -0.3 true cm \small $q \;$ [GeV]}
\psfrag{zq2}{\small $q ' \;$ [GeV]}
\psfig{file=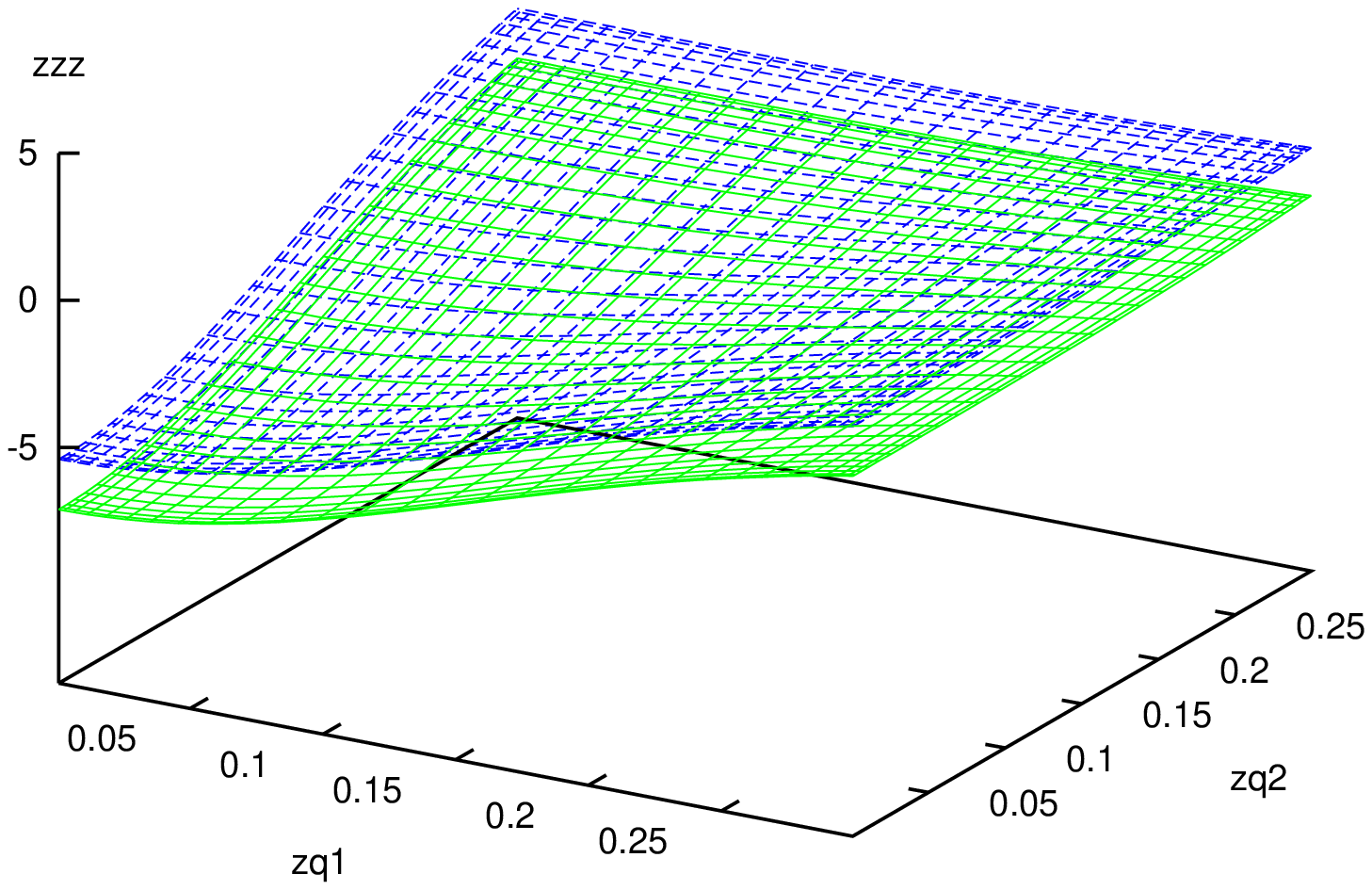,width=5in}}
\caption[potreg]{\label{fig10} Effective two--nucleon potential (solid
  green lines) in
comparison with the original potential (dashed blue lines) for momenta less 
than 300 MeV in the ${}^1 S_0$ channel.}
\end{figure} 
One can also perform a similar analysis for the two nucleon ${}^1 S_0$ channel,
which is expected to be even more troublesome than the ${}^3 S_1$ channel for 
the effective field theory approach since, as already stated before, 
the scattering lenth takes an unnatural large value. 
The ${}^1 S_0$ two--nucleon phase shift is
satisfactorily reproduced by choosing the following set of parameters for the
potential eq.(\ref{PotMT})~\cite{ETG}: $V_H=7.291$, $V_L=2.605$, $\mu_H=613.7$ MeV
and $\mu_L=305.9$ MeV. All parameters with the exception of $V_L$ remain the same
as for the ${}^3 S_1$ channel. $V_L$ is choosen somewhat smaller to
make the attraction weaker. With this parameters one finds for
the scattering length  $a=-23.6\,$fm, which is rather close to the
empirical value. The reason for this large value is the virtual bound
state of almost zero energy, which we, of course, also recover.
We have calculated the effective potential for $\Lambda=$300~MeV,
as shown in fig.~\ref{fig10}. Again the dominant effect of
integrating out the higher momenta seems to be given by an overall shift.
Using the same decomposition as in eq.(\ref{Vprime}), we have again expanded
${V '}_{\rm contact}$ into a string of local contact terms of increasing 
dimension, cf. eq.(\ref{const}). Keeping only the terms up to fourth order, we have 
fixed the  values of the constants $C_i$ by fitting  eq.~(\ref{potcont})
to $V '(q ',q)-V_{\rm light} (q ', q)$. The corresponding constants are
$C_0=10.62$~GeV${}^{-2}$, $C_2=-32.11$~GeV${}^{-4}$, $C_4=85.8$~GeV${}^{-6}$
and $C_4 '=117$~GeV${}^{-6}$, rather close to the constants from table~\ref{tab1}.
(The reason for this is that the term in the potential corresponding 
to the heavy meson exchange is not modified.)
Consequently, all conclusions about the size of the scale $\Lambda_{\rm scale}$ 
from the analysis of the ${}^3 S_1$ channel are valid in this case as well. 
The corresponding phase shifts are shown in fig.~\ref{fig11}.
\begin{figure}[htb]
\vspace{0.4cm}
\centerline{\psfrag{zzz}{\small $\delta \; [ \, {}^\circ \,]$}
\psfrag{zen}{\small $T_{\rm lab} \;$ [MeV]}
\psfig{file=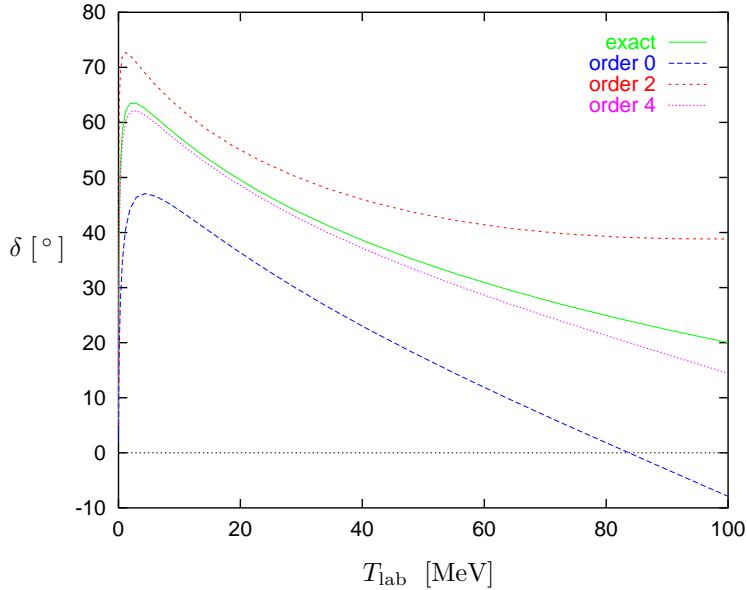,width=3.9in}}
\vspace{0.51cm}
\caption[phcont]{\label{fig11} Phase shifts from the
effective potential $V '$ (solid line) and
the truncated expansion eq.(\ref{potcont}) as a function of the kinetic energy
in the lab frame for ${}^1 S_0$ channel for  $\Lambda =$300~MeV.}
\end{figure}
Again as in the case of the ${}^3 S_1$ channel, one sees that the phase shift 
is described well up to about 100~MeV if the terms up--to--and--including fourth order 
in the expansion eq.(\ref{potcont}) are kept explicitly.
To further study the convergence of the contact term expansion eq.(\ref{potcont})
in the ${}^1 S_0$ channel, we have again calculated the expectation 
values of operators $V^{(0)}$, $V^{(2)}$ and $V^{(4)}$
using the scattering states for different energies.
As  exhibited in table~\ref{tab4} the values of the quantum averages
are slightly different from the ones for the ${}^1 S_0$ channel
and the expansion eq.(\ref{potcont}) is again convergent.
\begin{table}[h]
\centerline{\begin{tabular}{|c||c|c|c|c|} \hline
&$\langle \Psi | V^{(0)} | \Psi \rangle $ &
$\langle \Psi | V^{(2)} | \Psi \rangle $ &
$\langle \Psi | V^{(4)} | \Psi \rangle $  \\ \hline \hline
$E_{\rm lab} = 10$ MeV & 37.42 GeV${}^{-2}$ & -4.68 GeV${}^{-2}$ 
& 0.90 GeV${}^{-2}$ \\ \hline
$E_{\rm lab} = 50$ MeV & 11.72 GeV${}^{-2}$ & -2.56 GeV${}^{-2}$ 
& 0.68 GeV${}^{-2}$ \\ \hline
$E_{\rm lab} = 100$ MeV & 8.82 GeV${}^{-2}$ & -2.98 GeV${}^{-2}$ 
& 1.18 GeV${}^{-2}$ \\ \hline
\end{tabular}}
\caption{Quantum averages of the operators  $V^{(0)}$,
$V^{(2)}$ and $V^{(4)}$ for 
the scattering states with $\Lambda =300$ MeV in the ${}^1 S_0$
channel.}
\label{tab4}
\end{table}
However, one observes a much slower convergence for the scattering length, 
again because of its unnatural large value.
Keeping $V^{(0)}$, $V^{(0)}+ V^{(2)}$ and 
$V^{(0)} + V^{(2)} + V^{(4)}$ in the expansion for $V_{\rm contact}$ one 
finds $a=-9.0\,$fm, $a=-54.5\,$fm and $a=-21.2\,$fm, respectively. 
This is analogous to the situation  with the binding energy in the
${}^3 S_1$ channel, compare table~\ref{tab2} 
(some recent references dealing with the particular features of this partial
wave in the EFT approach are~\cite{kswold},\cite{ksw},\cite{silas}).
Certainly, one can achieve a considerably faster convergence with respect to the 
scattering length by  fine tuning  the constants $C_i$ (i.e. by fixing
them from a direct fit of the effective potential to the 
phase shift), as it was illustrated for the ${}^3 S_1$ channel.

\subsection{Coordinate space representation}

As discussed before, applying the method of unitary transformation turns
local operators in momentum space into non--local ones. Similar
effects happen in the coordinate space representation. To make this
transparent, we consider here the configuration space representation
of the effective potential $V'$. Denote by $V_l (p',p)$ the effective
momentum space potential for angular momentum $l$. The corresponding
r--space expression is obtained from
\beq
V_l (x',x) = \frac{2}{\pi} \int p^2 dp \, {p'}^2 dp' \, j_l (px) \,
V_l (p',p) \, j_l(p'x')~,
\eeq
with $j_l (y)$ the conventional $l^{th}$ spherical Bessel function.
The S--wave potential ($l = 0$) for $\Lambda = 400\,$MeV is shown in
the left panel of fig.~\ref{fig9}. It is, of course, symmetric under
the interchange of $x$ and $x'$, but looks very different from a local
potential.
\begin{figure}[htb]
\vspace{0.4cm}
\parbox{8cm}{
\centerline{\psfrag{zzz}{\hskip -0.8 true cm \small{$V_0 (x,x')$
[GeV${}^{4}$]}}
\psfrag{zq1}{\small $x$ [fm]}
\psfrag{zq2}{\small $x'$ [fm]}
\psfig{file=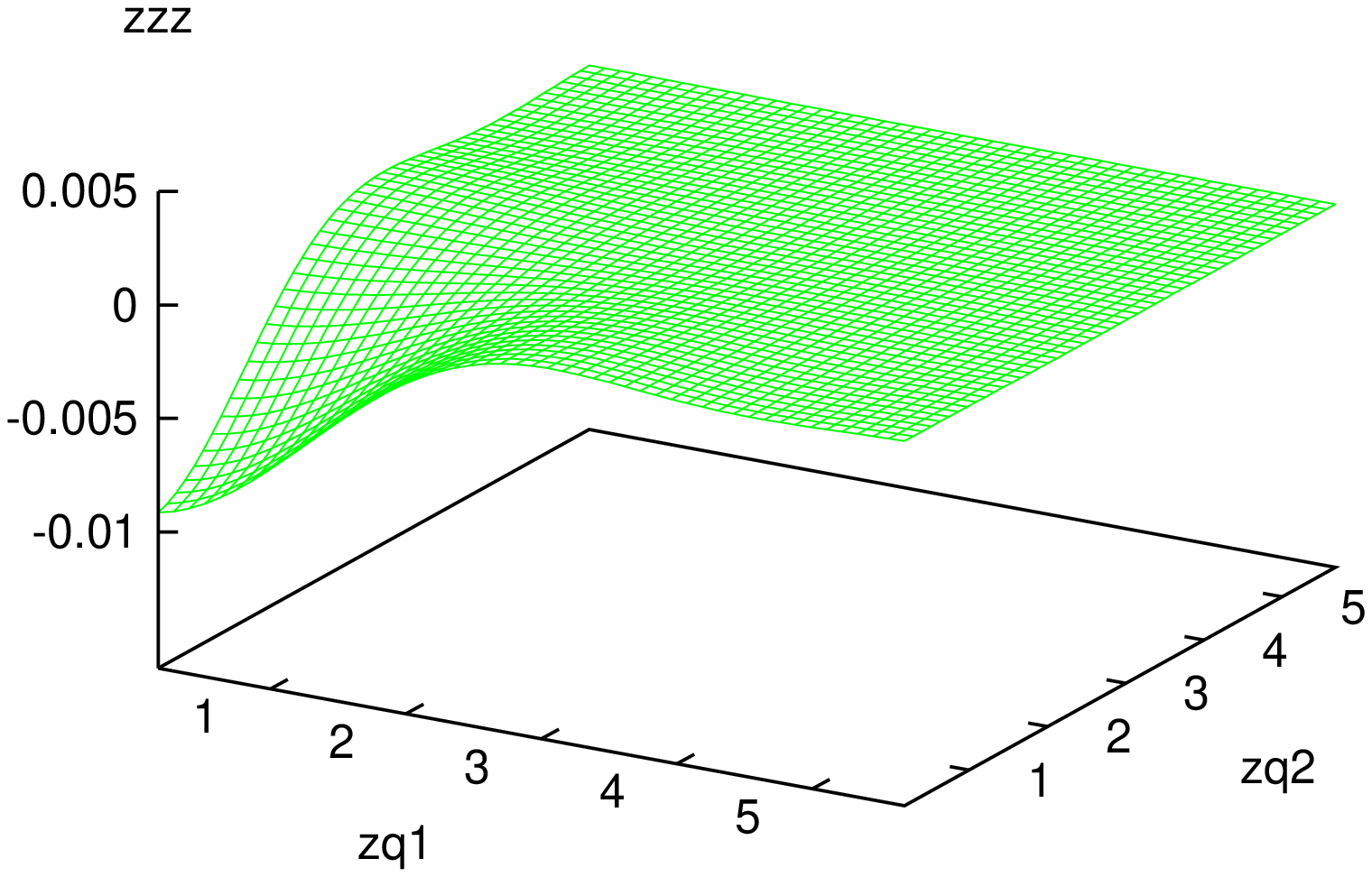,width=3.4in}}}
\hfill
\parbox{8cm}{
\centerline{\psfrag{zzz}{\hskip -0.8 true cm \small{$V_0 (x,x')$
[GeV${}^{4}$]}}
\psfrag{zq1}{\small $x$ [fm]}
\psfrag{zq2}{\small $x'$ [fm]}
\psfig{file=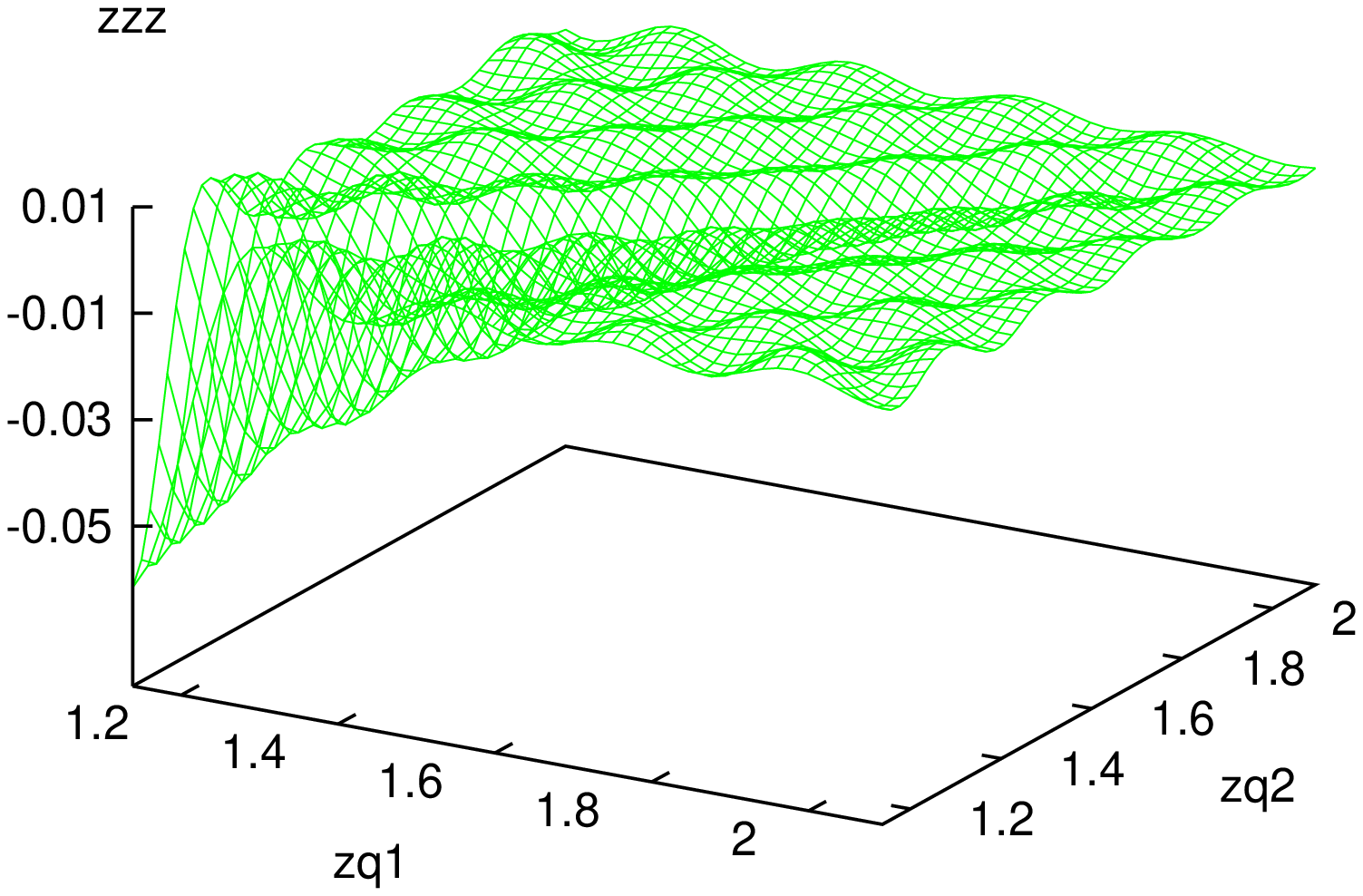,width=3.4in}}}
\vspace{0.51cm}
\caption[phcont]{\label{fig9} Coordinate space representation of
  the effective S--wave potential $V_0 (x,x')$ . Left panel: $\Lambda
  = 400\,$MeV and $0 \leq x,x' \leq 6\,$fm. Right panel: $\Lambda =
  5.5\,$GeV and $1 \leq x,x' \leq 2\,$fm.}
\end{figure}

Of course, if one  increases the
value of the the cut-off $\Lambda$,
a peak along the diagonal $x = x'$ resembling the delta function
should develop  and the superposition of the
two Yukawa potentials related to the light and heavy meson exchanges,
respectively,  should appear along the diagonal. This is indeed the
case  as demonstrated in the right panel of fig.~\ref{fig9} for
$\Lambda = 5.5\,$GeV. Note also that in this case, where $U \sim 1$,
the range of the potential is essentially given by the inverse of the
light meson mass. One can also construct the momentum and coordinate
representations of the deuteron wave function. For a momentum space
picture, we refer to ref.~\cite{egmplb}.

\section{Summary and conclusions}
\setcounter{equation}{0}

In this paper, we have shown how to construct an effective low energy
theory for nucleons based on the method of unitary transformations
starting from a realistic two--nucleon potential (in momentum space).
This unitary transformation can be parametrized by an operator $A$,
which obeys a nonlinear integral equation. This equation can be solved
numerically and any observable can then be calculated in the space
of small momenta only. To the best of our knowledge such an exact
momentum space projection has never been done before. While the method
is interesting {\it per} {\it se}, we have also made contact to chiral
perturbation theory (CHPT) approaches to the two--nucleon system by studying
a series of questions, which can be addressed unambigouosly within the
framework of our exact low momentum theory. Clearly, this should not
be considered a substitute for a realistic CHPT calculation but might
be used as a guide. The salient results of this investigation can be
summarized as follows:

\begin{enumerate}
\item[1)]We have demonstrated
that the theory projected onto  the  subspace of
momenta below a given momentum space cut--off $\Lambda$  leads to
exactly the same S--matrix as the  original theory in the
full (unrestricted) momentum space provided appropriate boundary conditions
for the scattering states are chosen. In particular, the components
of the transformed scattering states 
with initial momenta below the cut--off $\Lambda$
in the subspace of momenta above
the cut--off $\Lambda$ are strictly zero.
It is important to stress
that the exact projection leads to non--localities in momentum space.
\item[2)]Starting from an S--wave NN potential with an attractive
light ($\mu_L \simeq 300\,$MeV) and repulsive heavy meson exchange
($\mu_H \simeq 600\,$MeV), we have numerically solved without {\it any}
approximation the nonlinear equation for the operator $A$ and demonstrated that
the bound and scattering state spectrum of the effective and the
full theory agree exactly up to the cut--off $\Lambda$. In particular,
we have exactly one bound state with a binding energy of 2.23~MeV.
These results are independent of the value of the cut--off, which
was varied from 200~MeV to 5.5~GeV. We have argued that the most natural
choice is $\Lambda$ about 300~MeV.  The effective potential can differ
substantially from the original one (for values of $\Lambda$ on the
small side of the range mentioned before).
\item[3)]We have expanded the heavy meson exchange term in a string
of local operators with increasing dimension but kept the
light meson exchange explicitely. The corresponding coupling
constants accompanying these local operators,
which are monomials of even power in the momenta, can be determined precisely
from the exact solution. We have shown that they are of ``natural'' size,
i.e. of order one, with respect to the mass scale $\Lambda_{\rm scale}
= 600\,$MeV. We have also discussed the relation of this scale to the
mass of the heavy meson, which is integrated out, and the convergence
properties of such type of expansion. In particular,
to recover the binding energy within a few percent, one has
to retain terms of rather high order in this expansion, cf.
eq.(\ref{potcont}). This is to be expected due to the unnatural
smallness of this energy on any hadronic mass scale. The $^3S_1$
scattering phase shift can be well reproduced up to kinetic energies
$T_{\rm lab} \simeq 120\,$MeV with the first three terms in the contact
term expansion.  
\item[4)]Based on  the expanded heavy meson exchange term,
we have also determined the constants  $C_i$ directly from a fit
to the phase shifts. This is equivalent to the procedure performed in
an effective field theory approach. We could show that as long as one
does not include polynoms of order six (or higher), the resulting
values of these constants are close to their exact ones. Furthermore,
the binding energy is reproduced within 2\%. Including
dimension six terms, the fits become unstable. This can be traced back
to the fact that the contribution of such terms to the phase shifts
are very small (at low and moderate energies) and thus can not really be pinned
down.
\item[5)]We have also studied the  quantum averages of the expanded
potential in the bound and scattering states. For $\Lambda =
300\,$MeV, the expansion parameter is of the order of 1/2 and we find
fast convergence for the bound and the low--lying scattering states.
As expected,  for scattering states with higher energy, the
convergence becomes slower.
\item[6)]To study the $^1S_0$ channel, we had to slightly readjust the
parameters of the model potential. The phase shift can be well
reproduced with the terms up--to--and--including fourth order in the
contact term expansion of the heavy meson exchange. For the scattering
states, the quantum averages of the expanded potential show
convergence properties similar to the $^3S_1$ case. There is no bound
state in the $^1S_0$, but a virtual one just above
threshold. Therefore,
the pertinent scattering length is unnaturally large and it shows a 
similar slow convergence as does the binding energy in the $^3S_1$
channel.  
\item[7)]In the model space of small momenta only, one can also study
the non--localities in the coordinate space representation. We have
shown that for typical cut--off values, the effective potential
$V(x,x')$ is highly non--local and looks very different from the
original one. For very large values of the cut--off, one recovers the
original local potential.
\end{enumerate}
We hope that this study might be useful for derivation
of NN--forces based on chiral Lagrangians in the
low--momentum regime. It should
also provide new insights into a consistent and convergent treatment
of relativistic effects in few-- and many--nucleon systems.

\section*{Acknowledgements}

We would like to thank  Charlotte Elster, Henryk Wita{\l}a and
Hiroyuki Kamada for their help to solve various numerical problems
and for some useful numerical checks. We also thank Kenji Suzuki and
Ryoji Okamoto for some useful correspondence.

\bigskip\bigskip


\begin{thebibliography}{99}



\bibitem{weinberg}S.~Weinberg, Phys. Lett. B251 (1990) 288;
                  Nucl. Phys.  B363 (1991) 3.\vs
\bibitem{bira}C.~Ord\'{o}\~{n}ez, L.~Ray and U.~van Kolck, Phys. Rev. Lett.
              72 (1994) 1982; Phys. Rev. C53 (1996) 2086;
              U. van Kolck, Phys. Rev. C49  (1994) 2932.\vs
\bibitem{kswold} D.B. Kaplan, M.J. Savage and M.B. Wise,
  Nucl. Phys. B478 (1996) 629.\vs
\bibitem{ksw} D.B. Kaplan, M.J. Savage and M.B. Wise, Phys. Lett. B424
  (1998) 390;\\ {\tt nucl-th/9802075}.\vs
\bibitem{silas}S.R.~Beane, T.D.~Cohen and D.R.~Phillips,
Nucl. Phys. A632  (1998) 445.\vs
\bibitem{norb} N. Kaiser, R. Brockmann and W. Weise, Nucl. Phys. A625
(1997) 758;\\
N. Kaiser, S. Gerstendoerfer and W. Weise, {\tt nucl-th/9802071}.\vs
\bibitem{robi} M.R. Robilotta and C.A. da Rocha, Nucl. Phys. A615
(1997) 391;\\
J.-L. Ballot, M.R. Robilotta and C.A. da Rocha, Phys. Rev. C57 (1998)
1574.\vs
\bibitem{EGM} E. Epelbaoum, W. Gl\"ockle and Ulf-G. Mei{\ss}ner,
Nucl. Phys. A637 (1998) 107.\vs
\bibitem{kor} T.-S. Park, K. Kubodera, D.-P. Min amd M. Rho,
{\tt nucl-th/9807054}.\vs
\bibitem{silas2}S.R. Beane, {\tt nucl-th/9806070}.\vs
\bibitem{okubo}S.~Okubo, Prog.~Theor.~Phys.~12  (1954) 603.\vs
\bibitem{FST} N. Fukuda, K. Sawada and M. Taketani, Prog. Theor.
              Phys. 12 (1954) 156.\vs
\bibitem{SO} K. Suzuki and R. Okamoto, Prog. Theor. Phys. 70 (1983)
  439; Prog. Theor. Phys. 92 (1994) 1045.\vs
\bibitem{privet} K. Suzuki and R. Okamoto, private communication. \vs
\bibitem{sw} T.~H.~Schucan and H.~A.~Weidenm\"uller, Ann.~Phys. 73 (1972) 108. \vs
\bibitem{egmplb} E. Epelbaoum, W. Gl\"ockle and Ulf-G. Mei{\ss}ner,
{\tt nucl-th/9804005}, Phys. Lett. B439 (1998) 1.\vs
\bibitem{spli} W. Gl\"ockle, G. Hasberg and A. R. Neghabian,
Z. Phys. A305  (1982) 217. \vs
\bibitem{kuo} T.T.S. Kuo, Lecture Notes in Physics, Vol. 144, p.~248
(Springer Verlag, 1981).\vs
\bibitem{ETG} Ch.~Elster, J.H.~Thomas and W.~Gl\"ockle,
Few Body Syst. 24 (1998) 55.\vs
\bibitem{bira2} U. van Kolck, {\tt hep-ph/9711222}.\vs
\bibitem{hoeh}G. H\"ohler et al., Nucl. Phys. B114 (1976) 505.\vs
\bibitem{MMD}P. Mergell, Ulf-G. Mei{\ss}ner and D. Drechsel,
  Nucl. Phys. A596 (1996) 367.\vs

\end{thebibliography}
\end{document}